\documentclass[10pt,preprint]{emulateapj}

\usepackage{CJK}

\shorttitle{Empirical Isochrones for Low Mass Stars}
\shortauthors{Herczeg \& Hillenbrand}

\begin{document}
\begin{CJK*}{UTF8}{gbsn}

% Title.
\title{Empirical Isochrones for Low Mass Stars in Nearby Young Associations}

% Authors.
\author{Gregory J. Herczeg (沈雷歌)\altaffilmark{1} \& Lynne A. Hillenbrand\altaffilmark{2}}
\altaffiltext{1}{Kavli Institute for Astronomy and Astrophysics,
  Peking University, Yi He Yuan Lu 5, Haidian Qu, 100871 Beijing,
  People's Republic of China}
\altaffiltext{2}{Caltech, MC105-24, 1200 E. California Blvd., Pasadena, CA 91125, USA}

% Abstract
\begin{abstract}
Absolute ages of young stars are important for many issues in pre-main sequence stellar and circumstellar evolution
but are long recognized as difficult to derive and calibrate.  In this paper,
we use literature spectral types and photometry to construct empirical
isochrones in HR diagrams for low-mass stars and brown dwarfs in the
$\eta$ Cha, $\epsilon$ Cha, and TW Hya Associations and the $\beta$ Pic and Tuc-Hor Moving Groups.  
A successful theory of pre-main sequence evolution should match the shapes of the stellar loci for
these groups of young stars. However, when comparing the combined empirical
isochrones to isochrones predicted
from evolutionary models, discrepancies lead to a spectral type (mass)
dependence in stellar age estimates.  Improved prescriptions for
convection and boundary conditions in the latest models of pre-main
sequence models lead to a significantly improved correspondence
between empirical and model isochrones, with small offsets at low
temperatures that may be explained by observational uncertainties or by
model limitations.
Independent of model predictions, linear fits to combined stellar loci
of these regions provide a simple empirical method to order clusters
by luminosity with a reduced dependence on spectral type. Age estimates calculated from various sets of modern models
that reproduce Li depletion boundary ages of the $\beta$ Pic Moving Group also imply  a $\sim 4$ Myr age for the low mass members of
the Upper Sco OB Association, which is younger than the 11 Myr age that has been recently estimated for intermediate
mass members.
\end{abstract}
 
% Official ApJ keywords in alphabetical order.
% See http://www.journals.uchicago.edu/ApJ/information.html

% Citation commands:
%
% \citet{Nam98} for text citations, e.g. "Name (1998) says..."
% \citep{Nam98} for parenthetical citations, e.g. "is true (Name 1998)."
% \citep{Nam98,Alt01} to cite multiple refs, e.g. "(Name 1998; Alt 2001)"
% \citep{Nam98,Nam01} automatically merges, e.g. "(Name 1998; 2001)"
% \citet*{Nam98} or \citep* for the first citation of 3 author papers.
% \citet*{Nam98} or \citep* for the first citation of 3 author papers.
% \citep[e.g.,][]{Nam98} for leading text, e.g. "(e.g., Name 1998)"
% \citep[, for example]{Nam98} for trailing text.
% \citep[e.g.,][, for example]{Nam98} for leading and trailing text.
% \citealt{Nam98} is \citet{Nam98} without parentheses.
% \citealp{Nam98} is \citep{Nam98} without parentheses.
% \citeauthor{Nam98} only lists authors, e.g. "Name"
% \citeyear{Nam98} only lists the year, e.g. "1998"
% \citeyearpar{Nam98} only lists the year in parentheses, e.g. "(1998)"
% \citetext{\citealp{Nam98}; see also \citealp{Alt01}} embeds text, e.g.
%         "(Name 1998; see also Alt 2001)"
%
% Look up "natbib" on the web for a more detailed description.

%%%%%%%%%%%%%%%%%%%%%%%%%%%%%%%%%%%%%%%%%%%%%%%%%%%%%%%%%%%%%%

\section{INTRODUCTION}

Age estimates for pre-main sequence stars are critical in assessing the timescales
for the formation of stellar systems and the evolution of
molecular clouds.  The fraction of embedded
sources determines how long the dominant phase of stellar growth lasts
and how quickly the envelope disperses.  Similarly, when combined with disk fractions, age measurements place strict constraints on the time available
for giant planets to form.  Ages are also important in
quantifying how the stellar contraction leads to rotational spin-up and
changes in interior
structure for solar and intermediate mass stars, factors which combine to
determine the initial conditions for the evolution of magnetic activity for the duration of the
stellar lifetime.  Members of young stellar associations are
usually dated by comparing their temperatures and luminosities to
pre-main sequence tracks.  However, errors and biases in placing
stars on Hertzsprung-Russell (HR) diagrams lead to uncertainty in ages of different
clusters, while large luminosity spreads among members of individual clusters enhance
uncertainty regarding age dating of any individual star
\citep[see review by][]{Soderblom2014}.

Further uncertainties in pre-main sequence age estimates from HR diagrams are traced directly to the varied stellar
loci predicted from different stellar evolution models and to the lack of
consistency between predicted and observed stellar loci.  These differences have
been noted between low and intermediate mass stars
\citep[e.g.][]{Hillenbrand1997,Hillenbrand2008,Naylor2009} and among
low mass stars of different mass
\citep[e.g.][]{DaRio2010,Murphy2013,Kraus2014,Alcala2014,Herczeg2014,Malo2014b}.
Components of young binary systems frequently have different estimates for age despite an
expectation that they are co-eval \citep{Hartigan1994,Hartigan2003,Kraus2009,Gennaro2012}.
As a direct consequence, age estimates depend on the considered range in 
spectral type or stellar mass.  \citet{Hillenbrand2008}
found that in many regions G stars appears $\sim 2-5$ times older than K stars, depending on which 
pre-main sequence tracks were used.  \citet{Pecaut2012} quantified
these differences by measuring an age for the Upper Sco OB
Association of $11\pm2$ Myr, larger than the 5 Myr age measured
previously from pre-main sequence low mass stars \citep{Preibisch2002,Slesnick2006}.
Similarly, in both younger and older clusters, analyses of
main sequence turnoff and intermediate mass stars in color-magnitude
diagrams yield ages that are twice as large as those obtained from low
mass pre-main sequence stars \citep{Mayne2008,Naylor2009,Bell2013}. 

Nearby stellar associations with ages of 5--40 Myr 
 provide us with a laboratory to establish empirical isochrones, which
 can be used as a model-independent clock and as a benchmark for testing pre-main sequence models.
Within a single
association older than 10 Myr, the age spread expected from star formation
processes is small compared to the absolute
age.  Member stars have traveled far from any
remnant molecular material associated with their birth cloud and are
unaffected by interstellar extinction.  Most disks have already
dispersed by these ages, so observational uncertainties related to
accretion and circumstellar extinction are negligible for the vast
majority of stars in this age range.  In
recent years, the growing number of proper motions and
parallax distances have led to a substantial improvement in the
accounting of membership within young associations \citep[e.g.][]{Malo2013,Kraus2014}.
The HR diagrams of these nearby moving groups should consequently have only a
small age spread.  

At younger ages, placing stars on HR diagrams is more problematic because
accretion and extinction affects temperature and luminosity
measurements, and because the affect of any genuine age spread on
luminosities decreases (in dex)
with age.  Nevertheless, wide spreads in luminosities of 
members of young clusters have been interpreted as true age spreads.  For example, in the ONC
a 1.5 dex luminosity spread (Da Rio et al. 2010) led Reggiani et
al. (2011) to claim an age spread of 1.5--3.5 Myr, although some of
this spread may alternately be interpreted as a spread in radius from
different accretion histories \citep{Hartmann1997,Baraffe2009}.  Improved spectral
types and extinction measurements of two outliers in the ONC sample indicate that the luminosity
spread may be overestimated \citep{Manara2013}.  If we
consider $\sim 2$ Myr as an approximate age spread, then that would
lead to 0.16 dex and 0.08 dex spreads in luminosities of 5 Myr
old and 20 Myr old clusters of single stars (as assessed
for 3500 K stars from the Baraffe tracks, see \S 2.5).  These luminosity spreads are
consistent with the 0.05 to 0.15 dex errors in placing young stars in
the HR diagram \citep[see Table 1;][]{Preibisch2012,Soderblom2014}.

In this paper, we use literature spectral types and photometry to
construct empirical isochrones of luminosity versus temperature 
of diskless low mass stars in nearby associations and compare them to pre-main sequence
isochrones.  This approach is similar to the empirical isochrones
derived by \citet{Hillenbrand2009} and by 
\citet{Bell2012,Bell2013,Bell2014}, but is focused exclusively on
low mass stars in nearby young associations and moving groups and is
motivated by a desire for deriving simple formulations that
approximate empirical isochrones in HR diagrams.
In \S 2, we describe the literature data that is used to place stars on HR
diagrams and the pre-main sequence evolutionary tracks used for age
estimates.  
In \S 3 we develop empirical isochrones for a set of
benchmark young associations.  In \S 4, the empirical isochrones are 
compared to isochrones predicted from models of pre-main sequence
evolution to test evolutionary tracks and to obtain age estimates
for the young clusters for each model.  In \S 5, we discuss how the ages from
different tracks compare with independent age measurements.

\section{SAMPLE, DATA, and PRE-MAIN SEQUENCE EVOLUTION MODELS}
\subsection{Membership and Properties of Nearby Moving Groups}

The  $\beta$ Pic and Tuc-Hor  Moving Groups and the $\epsilon$ Cha, $\eta$ Cha, and TW Hya
Associations are used in this paper as benchmark clusters to measure empirical isochrones of
young regions.   The results obtained from these benchmark clusters
are then applied to the Upper Sco OB Association.
Searches for membership in existing all sky surveys, along with
analyses of proper motions and radial velocities, have
substantially improved the confidence in membership for these
associations.  Meanwhile, the
proliferation of  accurate parallax
measurements for the nearest groups has significantly reduced the
uncertainty in luminosities arising from errors in distance measurements.  When combined, these advances allow for
a robust analysis of the stellar loci of these clusters.

In some analysis below, the two
older associations (the $\beta$ Pic and Tuc-Hor Moving
Groups) and the three younger associations (TW Hya, $\eta$ Cha, and
$\epsilon$ Cha) are grouped together.  Gravity differences
between these clusters are described in \S 2.2-2.3, and the similarity in
ages amongst the older and younger groups will be established in \S 3.
The binary census is discussed in
this section, while the implementation of binaries is described in \S 2.4.

{\it Beta Pic Moving Group:}  Membership for the $\beta$ Pic moving group
is obtained from the analysis of 
\citet{Malo2013,Malo2014}, based on a $>90$\% probability
calculated from proper motion, radial velocity, photometry, and parallaxes
when available.  Spectral types are obtained from their compilation of
literature spectral types.  
Members with distances obtained from kinematic
modelling are not used in this paper because many targets are close
enough that the percentage uncertainties are too large$^3$.   Binarity was
accounted for based on the compilation of \citet{Malo2013}, which
relied heavily on \citet{Janson2012} for M-dwarfs.  A few objects and
binary information is added from \citet{Riedel2014}. 
\footnotetext[3]{The confirmed
members ($>90\%$ probability) with parallax distances occupy a tight stellar locus, with a
luminosity scatter of 0.14 dex around the best fit line.  Confirmed
members between 3300--5000 K with only kinematic distances have a
luminosity scatter (standard deviation about the mean linear fit) of
0.36 dex (0.25 dex if restricted to 3350--5000 K).}

{\it Tuc-Hor:}  The membership and spectral types for the Tuc-Hor
Association is obtained from \citet{Kraus2014}.  The distances used
here are from their kinematic analysis$^4$.  The spectral types used here
are obtained from their spectral analysis rather than the SEDs.
Binary
information is not included in our analysis of Tuc-Hor.
\footnotetext[4]{Unlike the case for the $\beta$ Pic Moving Group, the kinematic distances for Tuc-Hor produce a tight
stellar locus, perhaps because the average distance is larger so the
fractional uncertainty is smaller.  \citet{Kraus2014} calculate that a 5\% uncertainty in kinematic
  distance measurements applies to stars in their sample with a
  typical 5 mas yr$^{-1}$ error in proper motion.  This claim is supported by
  parallax measurements of five stars \citep{vanLeeuwen2007,Riedel2014} that
  differ by 4\% from the kinematic distances in Kraus et al.}

\begin{figure*}[*!t]
\epsscale{1.}
\plottwo{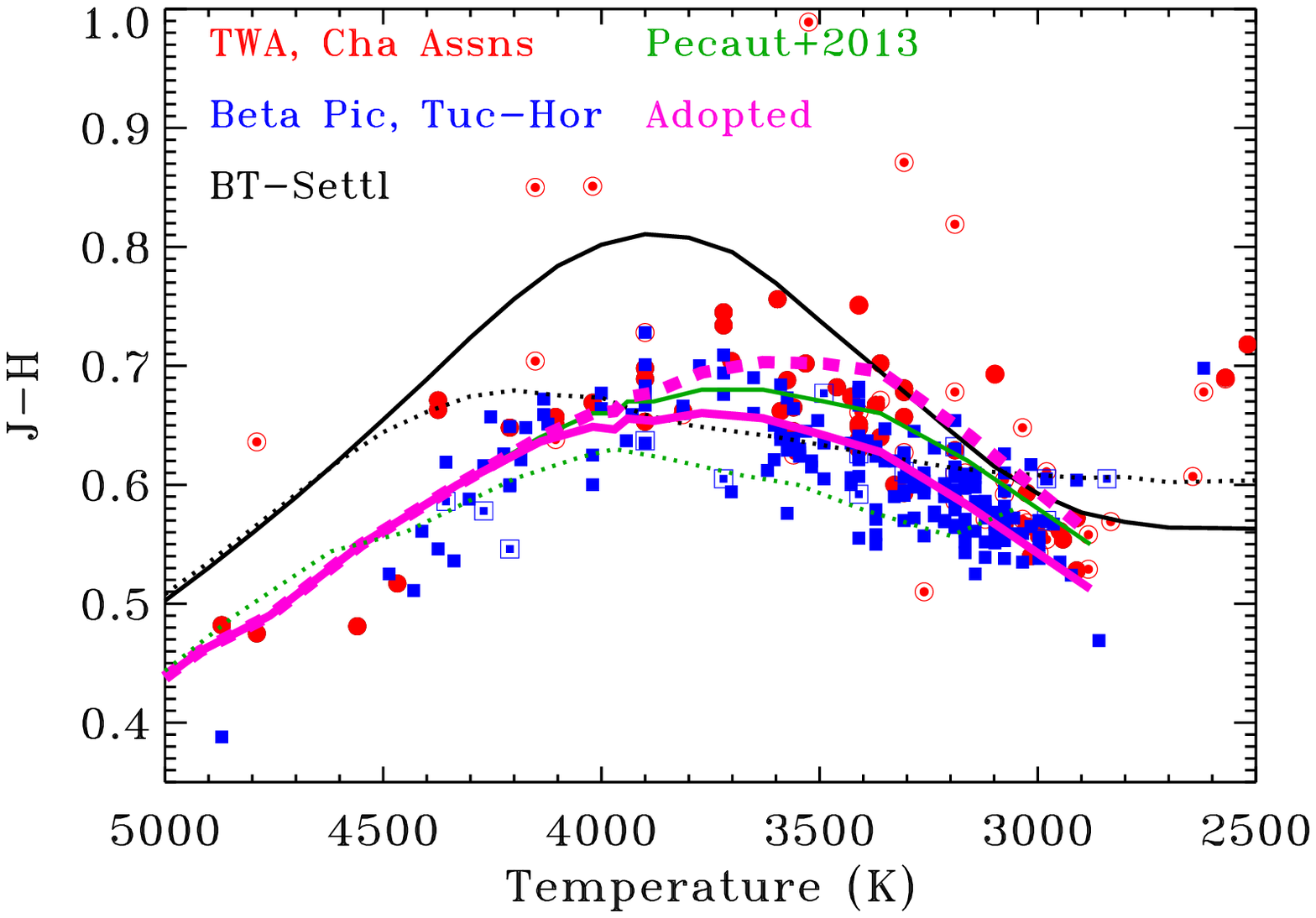}{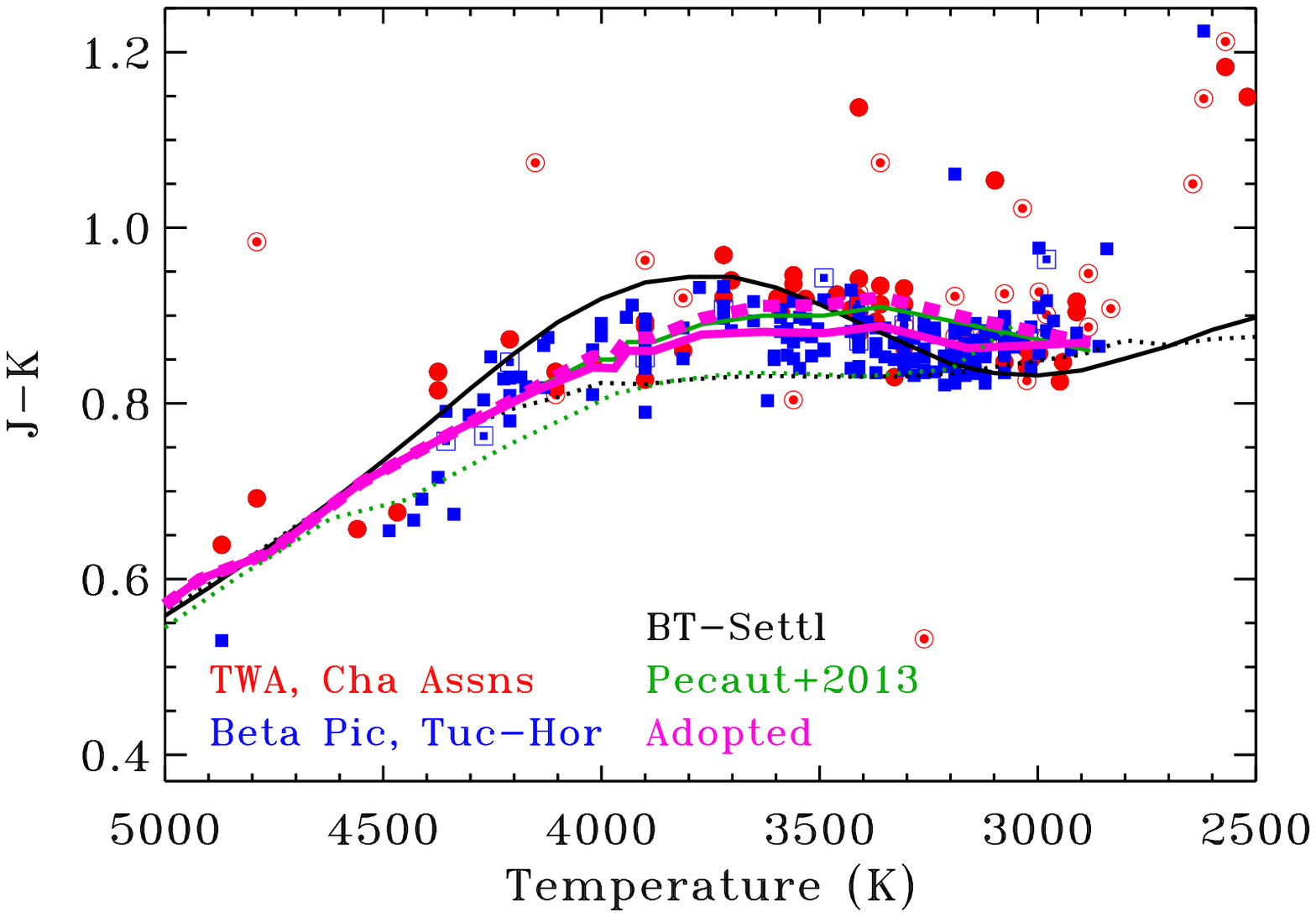}
\caption{The J--H and J--K colors for members of old (blue squares) and young
  (red circles, double circles indicate accretors) nearby
  associations, compared with colors from the BT-Settl models ($\log g=4.0$ is the solid line, $\log g=5.0$ is  the dotted
black line), the color from
Pecaut \& Mamajek (203, green line for young stars, dotted green line for
main sequence stars), and the color adopted here (purple solid line
for the older moving groups and purple dashed line for young associations).
The J-H colors of 
  $\beta$ Pic Moving Group and Tuc-Hor Moving Group members are offset
  by $\sim 0.05$ mag from the colors of TW Hya, $\eta$ Cha, and
  $\epsilon$ Cha Association members.  This measured gravity dependence is significantly less than predicted from the
BT-Settl models for M-dwarfs.  The J--H
color is adopted by approximating the J--H color intermediate
between the young and older clusters.}
\label{fig:jh}
\end{figure*}

{\it $\epsilon$ Cha Association:}  Membership, spectral types, and kinematic distances of  $\epsilon$ Cha members were obtained from \citet{Murphy2013}.  This
membership list is mostly consistent with but more complete than that
of \citet{Lopez2013}.  The targets listed as members in
\citet{Lopez2013} but excluded from \citet{Murphy2013} are VW Cha and
CM Cha, both of which have disks and would therefore not be included
in our analysis, and 2MASS J12074597-7816064, which was identified by
\citet{Fang2013} as a proper motion outlier.
The kinematic distances from \citet{Murphy2013} are adopted here and
have a median of 111 pc.
The known binaries are
accounted for following the description in \citet{Murphy2013}, though
the binary census is poor.

{\it $\eta$ Cha Cluster:}  Membership in the $\eta$ Cha cluster is
based primarily on the initial identification by \citet{Mamajek1999}
and on a follow-up census by \citet{Luhman2004}.  The spectral types
for those initial 18 members are obtained from \citet{Luhman2004}.
Five dispersed very
low mass stars were found and characterized by \citet{Murphy2010} and
are also included
here.  The additional members found by \citet{Lopez2013} based on proper motions
and the \citet{Covino1997} X-ray survey are excluded here because they are
far from the cluster core and may therefore be located at a different
distance.
The distance to all members is assumed to be 94 pc, the average
Hipparcos parallax distance of $\eta$ Cha and RS Cha (van Leeuwen
2007; see discussion in Murphy et al.~2010).
An accounting of spatially resolved binaries was obtained by
\citet{Brandeker2006} and is applied here.

{\it TW Hya Association:}  Membership in the TWA is based
on kinematic analyses \citep{Malo2013,Malo2014,Weinberger2013,Ducourant2014,Riedel2014}.  Spectral types of
most members and binarity are obtained from the compilation in
\citet{Herczeg2014}.  The new equal luminosity binary TWA member SCR
 1012-3124AB is also included here \citep{Riedel2014}. 
 Although parallax distances are preferred, TWA stars with kinematic
 distance measurements are included in our analyses.  For TWA 8A and 8B, a parallax distance
 distance of 44 pc is adopted here as the average of parallax
 distances measured by  \citet{Ducourant2014} and \citet{Riedel2014}.
 Three members of the HD 98800 quadruple system are included here with
 data from \citet{Laskar2009}, despite the presence of weak IR
 emission from HD 98800 N.

{\it Upper Sco OB Association:}  The stellar census and spectral types for the Upper Sco OB Association are
obtained from the compilation of members by \citet{Luhman2012}.  
Multiplicity is corrected based on several incomplete companion
searches \citep{Kraus2008,Kraus2012,Biller2011,Lafreniere2014}.  
The luminosities estimated for the Upper Sco OB Association members are corrected
for extinction (see \S 2.4).
 The average
extinction$^5$  of $A_V=0.46$ magfor low mass stars and $0.58$ mag for brown dwarfs is
intermediate between the average extinction of
$A_V=0.78$ mag for A stars, $A_V=0.59$ mag for G stars, and $A_V=0.35$ mag for F stars 
\citet{Pecaut2012}.  
 Stars with $A_V>2$ mag are
atypical for Upper Sco and are not included in the HR diagram analysis.  All stars in Upper Sco
are assumed to be located at a distance of 145 pc
\citep{deZeeuw1999}.  The cluster is 30 pc wide, and a depth of
similar length likely contributes to the large luminosity spread that
has been measured for Upper Sco \citep{Slesnick2008}.  
\footnotetext[5]{The extinction for both low mass stars and brown
  dwarfs is 0.66 mag if the photospheric J--K color is adopted from
  the older moving groups rather than the younger clusters.}

Stars without
any WISE W2 or W3 photometry (2 of 214 stars total between 3200--5000 K, 7
of 415 total brown dwarfs between
2800--3200 K) are included in the Upper Sco sample because their exclusion
could bias any results at low masses against faint objects.

\begin{table}[!b]
\caption{Approximate Uncertainties in Stellar Properties}
\label{tab:uncertainty}
\begin{tabular}{lcl}
Parameter & $T_{\rm eff}$ Range & Uncertainty\\
\hline
Temperature & $>4100$ K & 200 K or 2 SpT subclasses\\
Temperature & 3750--4100 K & 125 K or 1 SpT subclass\\
Temperature & 2800--3750 K & 75 K or 0.5 SpT subclasses\\
\hline
Luminosity & all & $~\sim 5\%$   parallax distance\\
Luminosity & all & 10-20\% kinematic distance\\
Luminosity & all & 2-5\% bolometric correction$^a$\\
\hline
\multicolumn{3}{l}{$^a$Or 0.02-0.05 in mag}\\
\end{tabular}
\end{table}%

\subsection{Colors of the Five Benchmark Clusters}

Photometry for members of the five benchmark clusters are
obtained from the 2MASS JHKs survey \citep{Skrutskie2006}, WISE W1,
W2, W3, W4 survey
\citep{Cutri2012}, the UCAC4 APASS all-sky BVgri survey
\citep{Zacharias2014}, and the GALEX NUV/FUV survey \citep{Bianchi2011}.   This collection of photometry is used in the
following subsections to identify and exclude disks, to ensure that
extinction is negligible (or correct luminosities for extinction, in the case of
Upper Sco members), and to calculate bolometric corrections (see also
Appendix A).  All objects are detected in JHK.  Some objects lack
photometry at other bands.

Broadband colors of objects in the younger TW Hya, $\eta$
Cha, and $\epsilon$ Cha Associations are mostly consistent with those from
the older 
$\beta$ Pic and Tuc-Hor Moving Groups, which all form a tight stellar
locus in infrared and optical colors (Figure~\ref{fig:jh} and
Appendix).   The contraction to the main sequence corresponds with a gradual
increase in gravity.  Just as different gravities between main
sequence and young stars yields color differences
\citep[e.g.][]{Luhman2003,Pecaut2013}, corresponding differences
should be present between 5 and 30 Myr old pre-main sequence stars.
The  J--H color is redder for the younger associations relative to the
older moving groups (Table~\ref{tab:bcor.tab} and right panel of Figure~\ref{fig:jh}), which is the
result of gravity differences and possible mismatches between SpT and
temperature between the two samples.  The J--K color is also $\sim
0.03$ mag redder
for early-to-mid M stars in younger associations.  Most empirical colors for the sample stars are also consistent with
the intrinsic colors for the 5--30 Myr stars as assessed by
\citet{Pecaut2013}.

\begin{figure}[!t]
\epsscale{1.}
\plotone{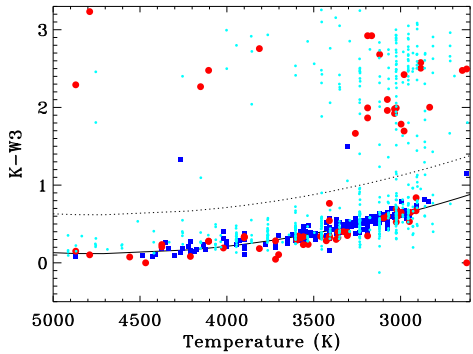}
\caption{The K-W3 color as a disk indicator.  The samples analyzed
  here are restricted to diskless stars, identified by a K-W3 color
  less than 0.5 mag above the best fit 2nd order polynomial (solid
  black line shows photospheric K-W3 color, dotted black line shows
  the cutoff for the presence of excess W3 emission).   The
  red circles are stars in the three younger associations, the blue
  squares are stars in the two Moving Groups, and the small cyan
  circles are stars in Upper Sco.}
\label{fig:diskcolor}
\end{figure}

The sample is restricted to diskless stars.  Disks are identified by
at least one of a K-W3 excess or K-W2 of 0.5 mag above the photospheric level, as determined
from the obseved stellar locus and from photospheric models
(Figure~\ref{fig:diskcolor}; see also, e.g., Luhman \& Mamajek~2012).  A few stars may retain
transition or debris disks that do not appear as a K-W2 or K-W3 excess, but
these disks are unlikely to affect the J-band brightness of the star.
The sources 2MASS J08014860-8058052, 2MASS J11550485-7919108, and 2MASS J02153328-5627175 fall below our disk criteria but are
excluded in our analysis because of excess emission in the WISE W4
band or the detection of
accretion in optical spectra \citep{Murphy2011,Murphy2013,Cutri2012}.

\begin{table}[!*hb]
\caption{Tempeartures and Bolometric Corrections versus Spectra Type}
\label{tab:bcor.tab}
\begin{tabular}{c|cc|cc|ccc}
       &  \multicolumn{2}{c}{$T_{\rm eff}$} & \multicolumn{2}{c}{BC$_J^a$} & \multicolumn{3}{c}{J-H}\\ 
SpT & HH14$^b$ & PM13$^c$ & Here & PM13$^c$ & Old$^d$ & Young$^d$ & PM13$^c$\\
\hline
F5&6600&6420&0.79&0.85&0.17&0.17&0.19\\
F8&6130&6100&0.95&0.96&0.22&0.22&0.23\\
G0&5930&6050&1.02&0.98&0.26&0.26&0.24\\
G2&5690&5870&1.10&1.03&0.30&0.30&0.27\\
G5&5430&5500&1.18&1.16&0.34&0.34&0.33\\
G8&5180&5210&1.26&1.25&0.40&0.40&0.39\\
K0&4870&5030&1.36&1.30&0.47&0.47&0.43\\
K2&4710&4760&1.41&1.40&0.50&0.50&0.49\\
K5&4210&4140&1.56&1.58&0.63&0.63&0.64\\
K7&4020&3970&1.62&1.63&0.66&0.66&0.66\\
M0&3900&3770&1.66&1.69&0.67&0.68&0.68\\
M1&3720&3630&1.73&1.74&0.67&0.70&0.68\\
M2&3560&3490&1.78&1.80&0.65&0.70&0.67\\
M3&3410&3360&1.84&1.84&0.63&0.70&0.66\\
M4&3190&3160&1.93&1.91&0.58&0.65&0.62\\
M5&2980&2880&1.99&2.01&0.53&0.59&0.55\\
M6&2860&&2.03&&0.50&0.55&\\
M7&2770&&2.06&&0.48&0.53&\\
\hline
\multicolumn{8}{l}{$^a$Following \citet{Pecaut2013}, the zero
    point for the }\\
\multicolumn{8}{l}{~~~J-band magnitude is
    $3.129\times10^{-10}$ erg cm$^{-2}$ s$^{-1}$ \AA$^{-1}$ and}\\
\multicolumn{8}{l}{~~~for the absolute
    magnitude is
    $3.055\times10^{35}$ erg cm$^{-2}$ s$^{-1}$}\\
\multicolumn{8}{l}{$^b$HH14 refers to \citet{Herczeg2014}}\\
\multicolumn{8}{l}{$^c$PM13 refers to the 5--30 Myr stars in Table 6}\\
\multicolumn{8}{l}{~~~of \citet{Pecaut2013}}\\
\multicolumn{8}{l}{$^d$J--H colors given for the median of diskless stars
  from the}\\
 \multicolumn{8}{l}{~~~two old moving groups and three young associations}\\
 \end{tabular}
\end{table}

\subsection{Bolometric Corrections and Temperature Scale}

The luminosities of stars in the five young, benchmark associations
are calculated by applying bolometric corrections to 2MASS J-band
magnitudes \citep{Skrutskie2006}.
Any bulk offset between the actual and implemented bolometric correction would
lead to an offset in luminosity and therefore age, while any temperature dependence
in bolometric correction offset would change the relative luminosities/ages of
stars in the same association but
with different temperatures.  
These systematic uncertainties introduce problems when comparing different methods and when
comparing observations to models. 
In this subsection, we calculate 
bolometric corrections and subsequently assess temperature dependence in the
uncertainties by comparing results from temperature and bolometric
corrections derived here to those of \citet{Pecaut2013}.  The Appendix 
describes the color corrections and constants that are used here to
calculate fluxes and stellar luminosities.

\begin{figure*}[!t]
\epsscale{1.}
\plottwo{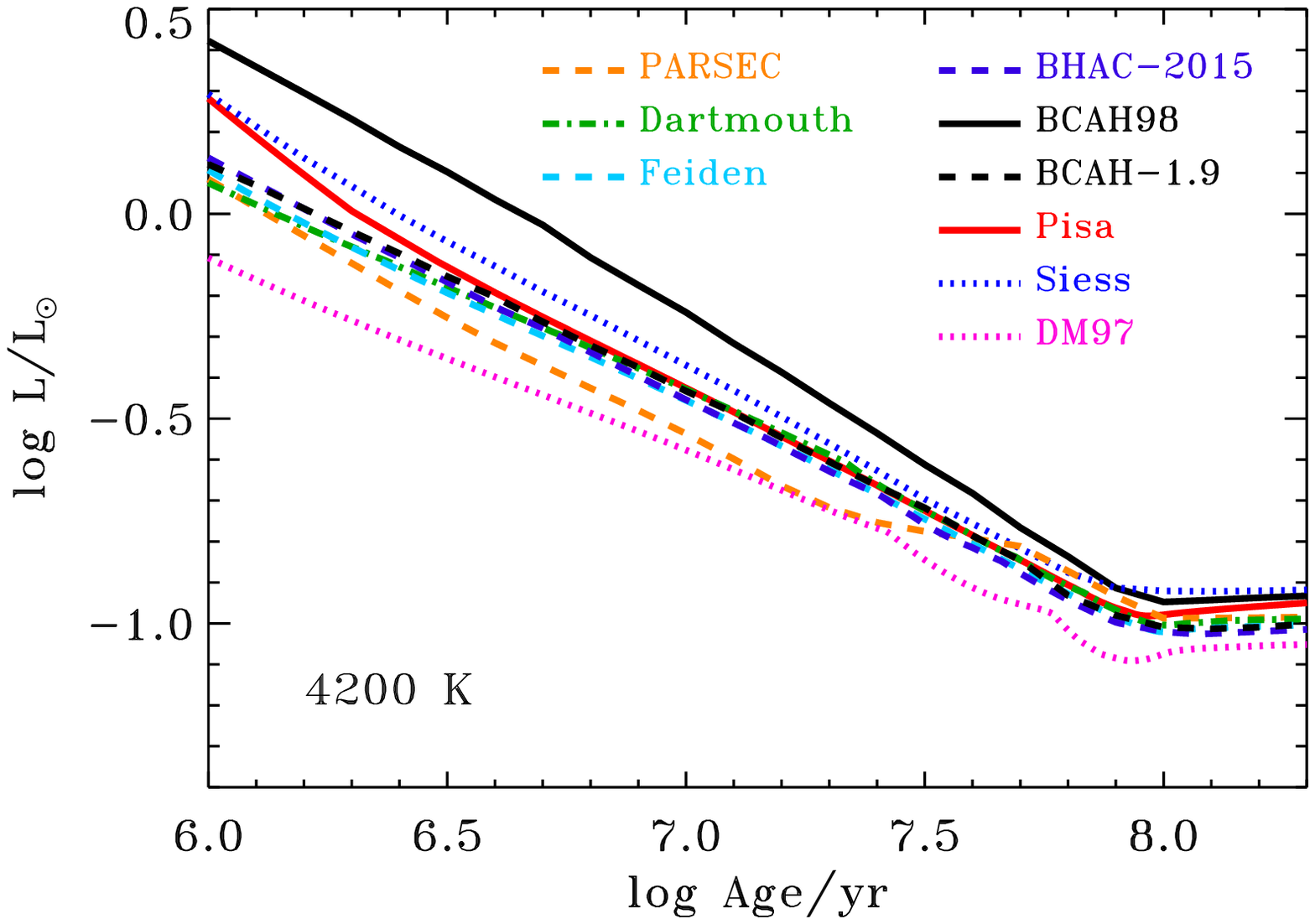}{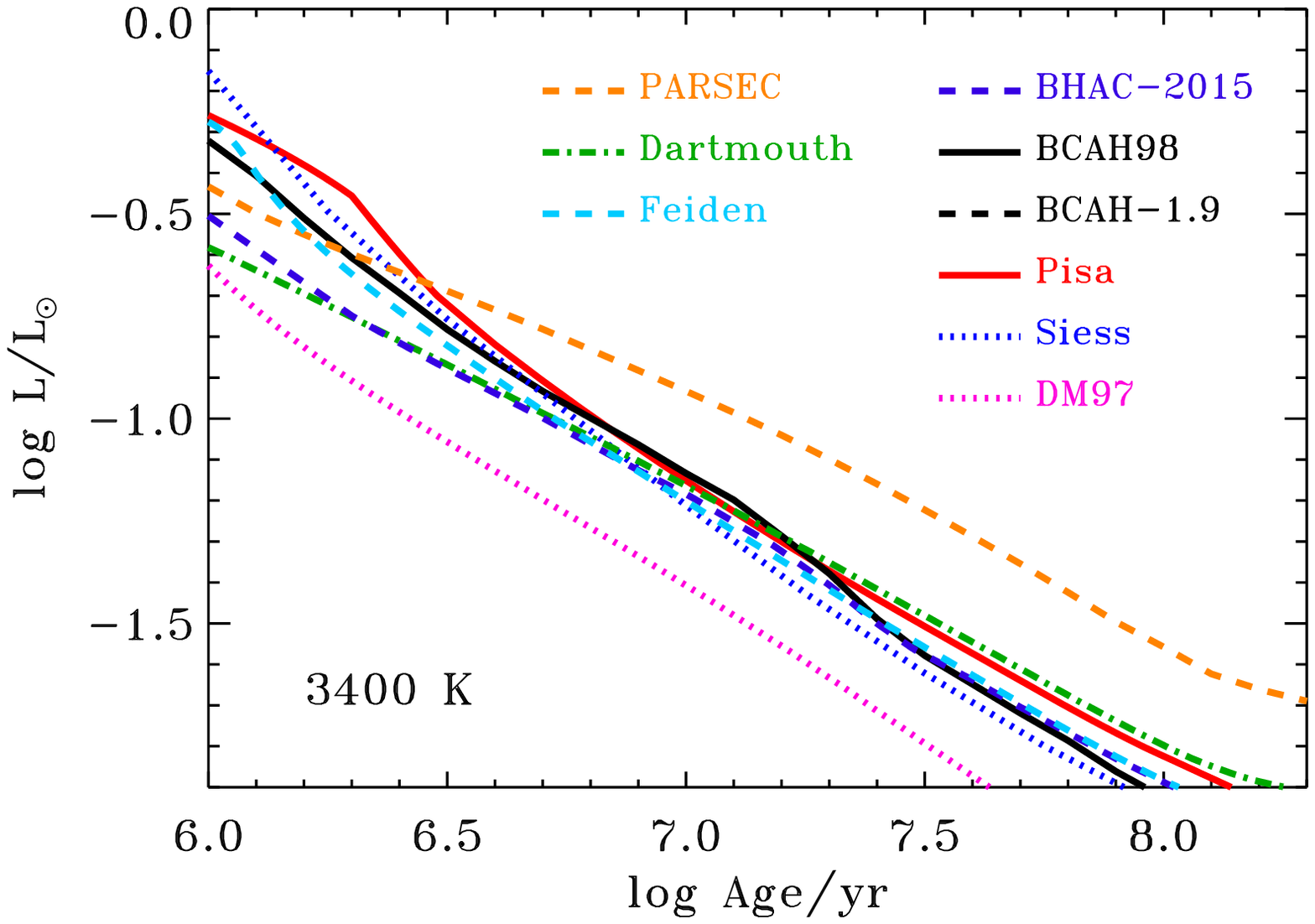}
\plottwo{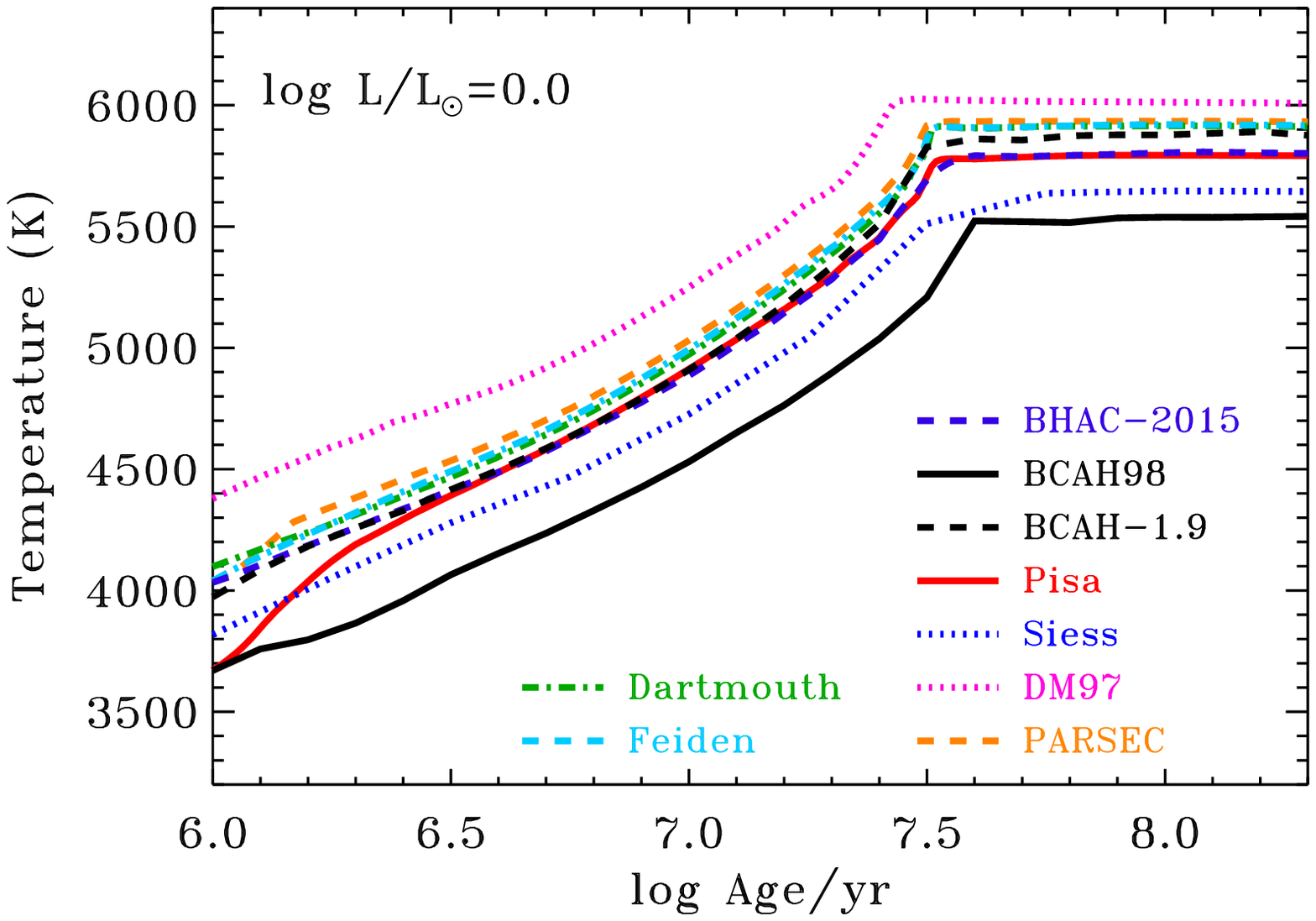}{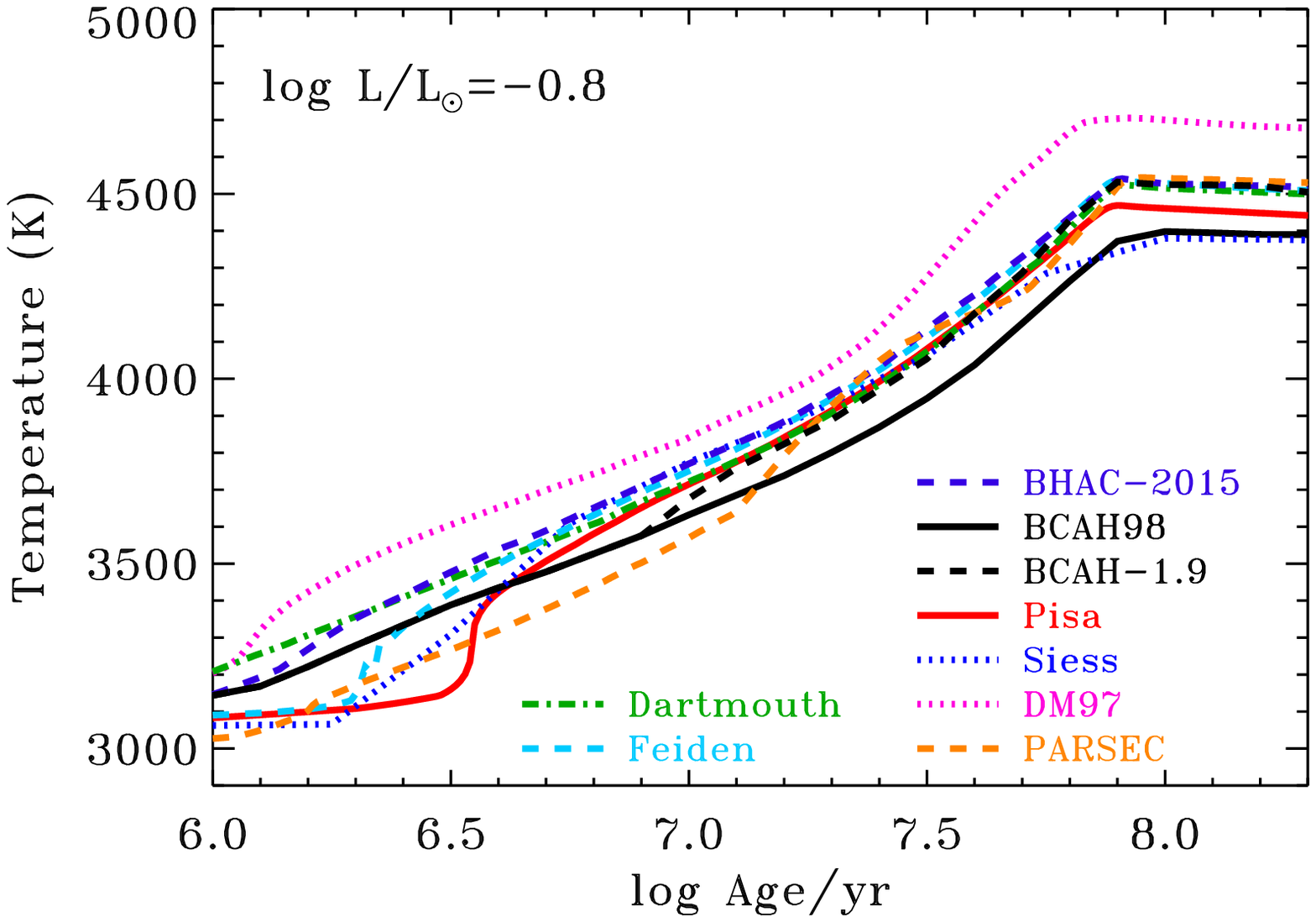}
\caption{Predictions from different pre-main sequence evolutionary
  tracks.  The top panels show how luminosity is converted to age for
  4200 K stars (roughly K5 spectral type, left panel) or 3400 K stars (roughly M3
  spectral type, right panel).  The bottom panels illustrate the
  temperature predictions at fixed luminosity.  
  The effective temperatures are much higher in the DM97 models
  and, for $L=L_\odot$, lower in the BCAH98 models.}
\label{fig:age3900k}
\end{figure*}

Our bolometric corrections are calculated 
from the BT-Settl model atmospheres (version CIFITS2011bc, which use
abundances from Caffau et al.~2011) with $\log g=4.0$
\citep{Allard2012}.  Minor offsets relative to the J-band flux are applied to several spectral
bands to account for differences between synthetic
and observed photometry (see Figure~\ref{fig:colors}).
Spectral types are converted to temperature from the conversion scale
of \citet{Herczeg2014}, which were calculated by comparing low resolution optical blue-red spectra to
2011 BT-Settl models.
The model spectra and observed photometry
are compared in magnitude space.  The synthetic photometry is
calculated by convolving the BT-Settl model spectra in photon units
with filter transmission curves (see Appendix A).  The
empirical colors are
obtained from the photometry described in \S 2.2.  The primary
uncertainties in these empirical corrections are introduced
by the lack of photometric coverage in the z and Y bands and in near-IR regions with poor
telluric transmission, which are located between the zYJHK filters.

Table~\ref{tab:bcor.tab} compares our spectral type-temperature
conversion and bolometric corrections to the 5--30 Myr stars  of \citet{Pecaut2013}.
In general, temperatures in \citet{Pecaut2013} are 100--200 K cooler
than those presented here.  These errors are roughly assessed from the
uncertainty in assigning spectral types from low
resolution optical spectra.
\citet{Pecaut2013} also include K8 and K9
spectral types.  Their SpT-temperature conversion extend only to M5, so
results for brown dwarfs (\S 3.2) are limited to the conversions of
\citet{Herczeg2014}. 
When evaluated at the same temperature, the bolometric corrections measured directly from the
BT-Settl models with the \citet{Caffau2011} abundances reproduce the \citet{Pecaut2013} bolometric
corrections to within 1\% at all spectral types, even though
\citet{Pecaut2013} used models with abundances from \citet{Asplund2009}. 
Differences between our bolometric corrections and those of Pecaut et
al.~are introduced by differences in the SpT-temperature conversion
and by our tweaking of the BT-Settl models to reproduce the measured colors,
relative to J.  The bolometric correction versus
spectral type is also different because of the different spectral
type-temperature conversion scale.  As a consequence, when compared by temperature, bolometric corrections are similar
for M-dwarfs and $\sim 2$\% smaller here for K-dwarfs.
The practical result of these differences is that  temperatures are $\sim 50-150$ K
higher and luminosities are $\sim1.5-4.5\%$ brighter relative to those
calculated from \citet{Pecaut2013} bolometric correction and
temperature scales.

The
analysis of stellar temperatures and luminosities for 3200--5000 K
stars (\S 3.1) is repeated
for the \citet{Pecaut2013} SpT-temperature conversions and bolometric
corrections.   Both \citet{Pecaut2013} and \citet{Herczeg2014} base
their SpT-temperature conversions on the BT-Settl models, although
\citet{Tottle2015} find discrepancies between these models and near-IR
colors at both early and late-M spectral types. 
While the SpT-temperature conversion is still a
significant source of uncertainty, the main results of this paper are
robust to systematic uncertainties in these scales.

\subsection{HR Diagram Methodology}

Stellar luminosities are calculated from 2MASS J-band brightness,
literature spectral types, and  stellar distances calculated from parallax measurements or
from association kinematics.    
The J-band is used for luminosity calculations because of the uniform reliability of the
2MASS survey, and because the J-band is near the peak of the spectral energy
distribution for late type stars, which minimizes the uncertainty in bolometric
correction caused by spectral type errors.  The stars are diskless, so accretion and
disk extinction do not affect the observed  J-band photometry.
Stars in each association are assumed to be co-eval.

Table~\ref{tab:uncertainty} lists the
uncertainties applied here for luminosity and
temperature estimates. 
Luminosity uncertainties are calculated from an assumed 5\% uncertainty from
a combination of photometric and bolometric correction errors and from
distance uncertainties, typically $<5$\% for parallax distances
(5\% distance uncertainty is assessed for all $\eta$ Cha Association
members), 5\% for kinematic distances for Tuc-Hor members (see
footnote 3), and 10\% for
kinematic distances of other regions.
Bolometric corrections for the J-band are 
described in detail in the next section and are more uncertain for cooler stars with large
opacities in molecules.  The depth of Upper Sco adds uncertainty in
the distance and therefore luminosity for individual members, which
contributes to its luminosity spread.  Temperature uncertainties
are based on estimated errors and methodological differences in
spectral typing (see discussion in Herczeg \& Hillenbrand 2014).
Systematic differences in spectral typing between different studies
will also affect the HR diagrams constructed in this paper.

In multiple systems, all members with resolved spectral types are
included in this analysis.  Unresolved spectral types are included if
the stars are spatially resolved in J, H, or K-bands.  In these cases,
the primary is assumed to have the system spectral type.  The J-band
magnitude of the primary is calculated by assuming that the brightness
difference is constant in the near-IR (the J--H and J--K colors
roughly constant for M-dwarfs).  Optically resolved or double-lined spectroscopic binaries
are included if both stars are similar brightness.  Single-lined binaries are excluded.

\begin{figure}[!t]
\epsscale{1.}
\plotone{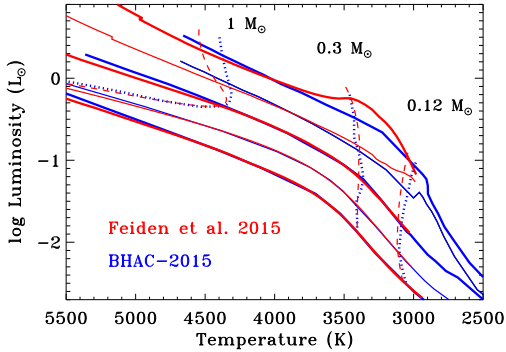}
\caption{A comparison of the new Feiden
  (red) and BHAC-2015 (blue) tracks with an improved
  treatment of convection.  The solid lines show isochrones at $\log$ age/yr=6.0, 6.5,
  7.0, 7.5, and 8.0, with 6.0, 7.0, and 8.0 in bold.  The vertical
  lines show the early evolution of $0.12$, $0.3$, and $1.0$ $M_\odot$
  stars from $\log$ age/yr=5.7 to 8.0.  Although the physics
  implemented into these modern tracks are different, they produce
  evolutionary tracks and isochrones that are extraordinarily
  similar.}
\label{fig:baraffefeiden}
\end{figure}

Extinction is negligible and assumed to be negligible for all diskless
stars in the five benchmark associations (see colors in \S 2.2 and the
Appendix).  Stars in the 
younger associations have a slightly bluer g-J color, which is the
opposite of expectations if extinction were important and supports the
assumption that extinction is negligible.  
The four stars in $\epsilon$ Cha with $A_V>1.5$ mag, as
calculated by \citet{Murphy2013}, all have disks, which may either
produce extinction or may alter the near-IR colors sufficiently to
have led to extinction in their calculations.

However, extinction
affects luminosities of stars in the Upper Sco OB Association.   A
color excess E(J-K) is calculated for each star here from by measuring
the difference between the observed J--K color the J--K color adopted
for the young associations, as shown as the dashed purple line in Figure~\ref{fig:jh}.   The
extinction $A_J$ is then calculated and corrected for by assuming a
total-to-selective extinction of $A_J=0.282 A_V$,
and $A_J=1.66 E(J-K)$ from \citet{Rieke1985}.  If the
total-to-selective extinction $R_V=4$ instead of 3.1, the $A_J$ values
would increase by 0.07\%, as calculated from the extinction curves of \citet{Weingartner2001}.

Young K and M stars without
disks have extinctions accurate to $\sim 0.2$ mag, although the
precise error in extinction depends on the error in spectral type.
Stars with extinctions $A_V>2$ mag are excluded from our analysis.
Negative values for extinction are retained and applied for statistical
purposes.

\begin{figure*}[!t]
\epsscale{1.}
\plottwo{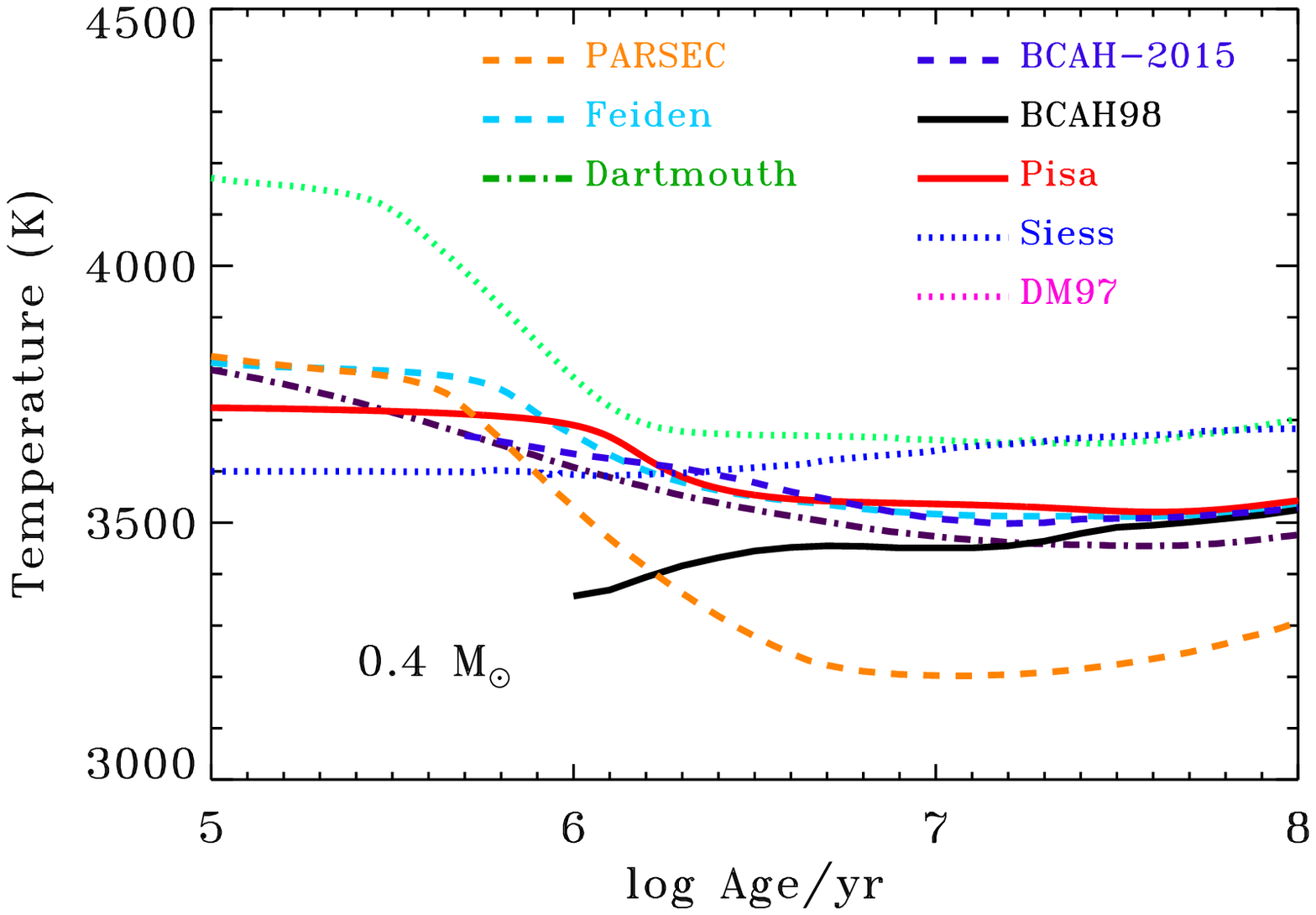}{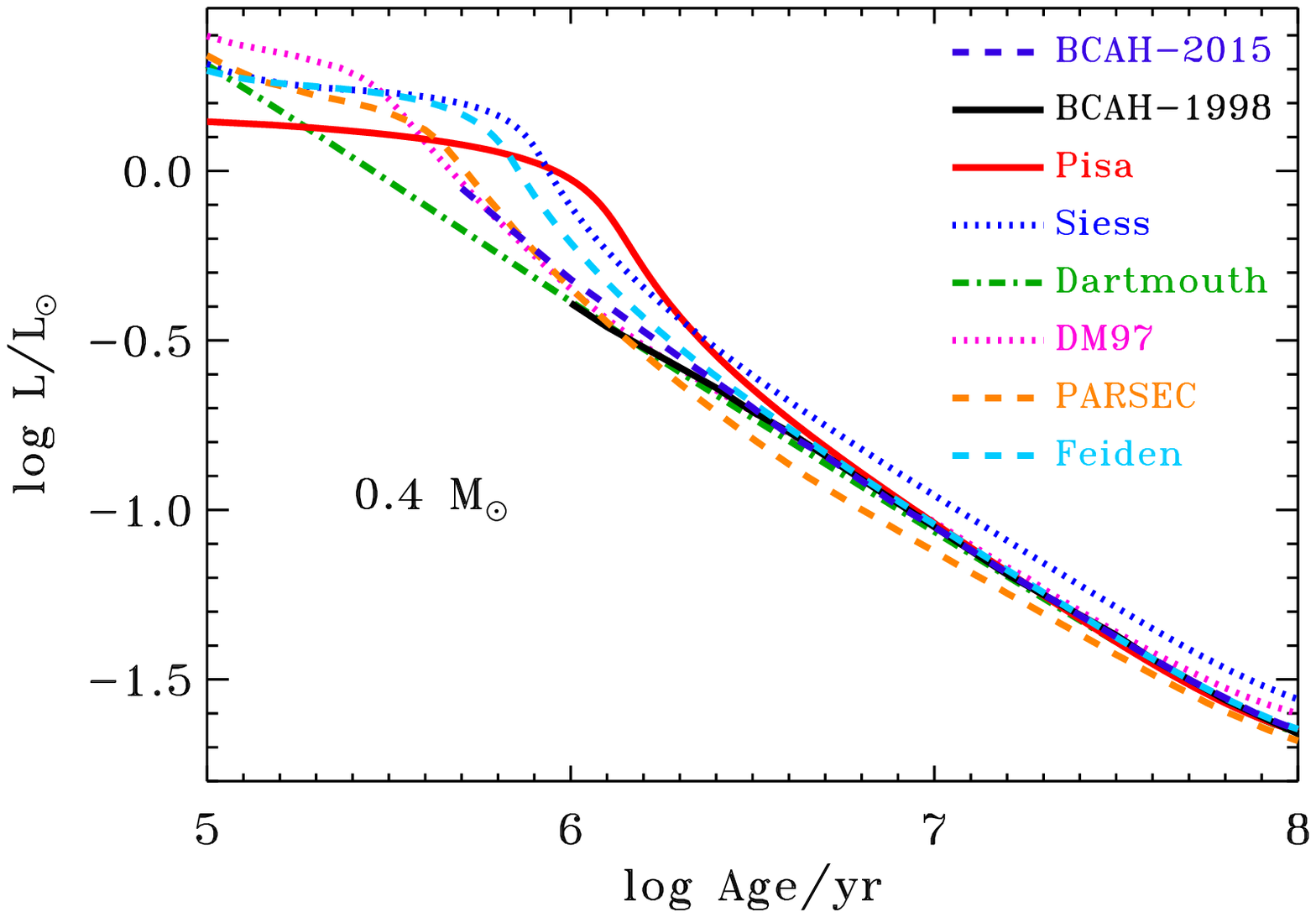}
\caption{The evolution of temperature and luminosity for a 0.4
  M$_\odot$ star, as predicted by different pre-main sequence models.  The
  temperature predictions are disparate at the 10-15\% level among
  models at all ages.  However, the
  luminosity evolution are similar after 2 Myr
for all models of the same mass, despite the different effective
temperatures.   The BHAC-2015
  tracks follow the Feiden tracks and are not shown here.}
\label{fig:massconv}
\end{figure*}

\subsection{Models of pre-main sequence evolution}

The empirical isochrones derived in \S 3 are compared to pre-main
sequence isochrones in \S 4.  \citet{Stassun2014} summarizes
currently available pre-main sequence evolutionary tracks.  The
evolutionary models used here, D'Antona \& Mazzitelli (1997, hereafter
DM97), Siess et al.~(2000, hereafter Siess), Baraffe et al.~(1998,
2003, hereafter BCAH98), Dotter et al.~(2008, hereafter Dartmouth),
Tognelli et al.~(2011, hereafter Pisa), and PARSECv1.2s (Bressan
et al.~2012, Chen et al. 2014; herafter PARSEC, also frequently
referred to as Padova), are selected based 
on their availability and prevalence in the
pre-main sequence literature. 
The BCAH98 models have
mixing lengths $\alpha=1.0$ (BCAH98) or mixing length
$\alpha=1.9$ (hereafter BCAH98-1.9).
The PARSEC models are selected for their artificial
tweaks to the $T-\tau$ relationship that reproduce the
color-magnitude diagrams of
very low mass stars in Praesepe and M67.  For stars below 4000 K,
stars of the same age and mass are much cooler and are only slightly
less luminous (because of larger radii) in the PARSECv1.2s models
relative to the PARSECv1.1 models \citep{Bressan2012}.  The PARSECv1.1
models are used only in Table~\ref{tab:slopes.tab} and are not 
discussed further.

In addition to these sets of models, we include analysis of newly
available tracks calculated by Baraffe et
al.~(2015, hereafter BHAC-2015) using the BT-Settl atmospheric models
and a prescription of
convection based on 2D/3D models, which is an improvement from the mixing length
prescription for convection.  We also include new 
Dartmouth tracks, which improve the thermal structure in the
atmosphere by calculating fitting to large optical depths (Feiden
et al.~in prep., private communication, hereafter Feiden).  The Feiden
models use the \citet{Grevesse1998} abundances with PHOENIX AMES-COND atmospheres \citep{Hauschildt1999a,Hauschildt1999b}.
\citet{Stassun2014} include preliminary descriptions of the Feiden
tracks, while \citet{Malo2014b} use a magnetic version of the Feiden tracks.
The BHAC-2015 and Feiden models
predict very similar temperature/luminosity evolution for low mass stars
and brown dwarfs, which then yields very similar isochrones (Figure~\ref{fig:baraffefeiden}).

All model tracks use solar abundances$^6$, which
is consistent with abundance measurements of young stars in the
solar neighborhood \citep[e.g.][]{Dorazi2011}.  Only the BCAH98
 and the DM97 pre-main sequence
tracks are available for masses $<0.1$ $M_\odot$, while the
Pisa models are only available for masses $>0.2$ M$_\odot$.  The
BCAH98 models with $\alpha=1.0$ are available for stars less massive than $<1.4$ $M_\odot$
while only models between $0.6-1.4$ $M_\odot$ have been calculated
with $\alpha=1.9$.  The PARSEC models are only calculated to $0.1$
M$_\odot$, but extend to low temperatures (2800 K at 1 Myr, 2500 K at
10 Myr) because the tweaks to the $T-\tau$ relationship reduce the
effective temperature of very low mass stars.
\footnotetext[6]{Changes in solar abundance measurements means that
  these models may use different abundances and are therefore not
  directly comparable.}

Figure~\ref{fig:age3900k} shows
the conversion to age at fixed luminosity and at fixed
temperature for the different pre-main sequence evolutionary models.  
 The DM97 model tracks are significant outliers \citep[see also, e.g.][]{Hillenbrand1997,Bell2012,Andrews2013}.  Their
photospheric temperatures are hotter for stars of the same mass, so
the corresponding mass of K and M stars is much lower than obtained from
other evolutionary tracks.  The measured
luminosity for a star of that mass is therefore high, leading to a young
age.  At ages $<3$ Myr, the Dartmouth tracks
also yield ages that are younger than the other tracks.
In contrast, the BCAH98 tracks assessed for 4200 K stars are outliers in
yielding older ages.

All models predict similar luminosities for stars of the same mass
for ages older than 2 Myr, despite different temperatures (Fig.~\ref{fig:massconv}).  With accurate age dating of young clusters, masses may
be inferred from the luminosity alone and with consistent results
between different model
tracks.

\section{EMPIRICAL ISOCHRONES FOR THE NEARBY YOUNG MOVING GROUPS}

With an accurate theory of pre-main sequence evolution, 
the estimated luminosity of a star at a given estimated temperature could be directly
converted to an age.  However, scatter in $L$ and $T_{\rm eff}$ among
young cluster members provokes concerns regarding the reliability of 
age estimates for
single stars.  Moreover, differences in the slope between observed
and model isochrones lead to ages that depend on spectral type.
Our approach here focuses on comparing the empirical stellar locus of different
associations and interpreting the offsets as relative age
differences.  Using the stellar loci constructed for young associations, we first focus on empirical trends in $\log
L$ versus $T_{\rm eff}$ (this section) before conducting detailed
comparisons to models (\S 4).

\begin{figure*}[!t]
\epsscale{1.}
\plotone{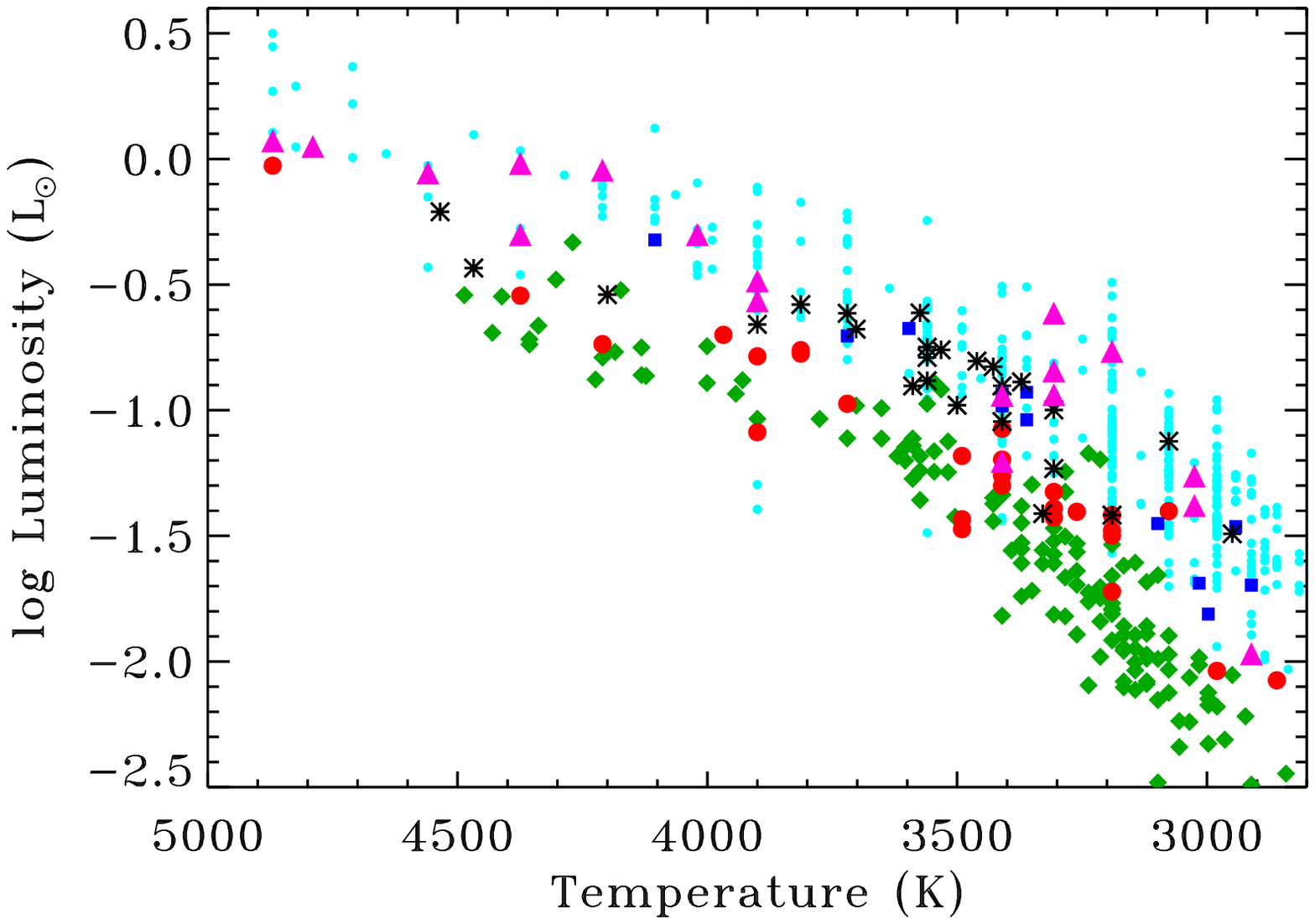}
\plottwo{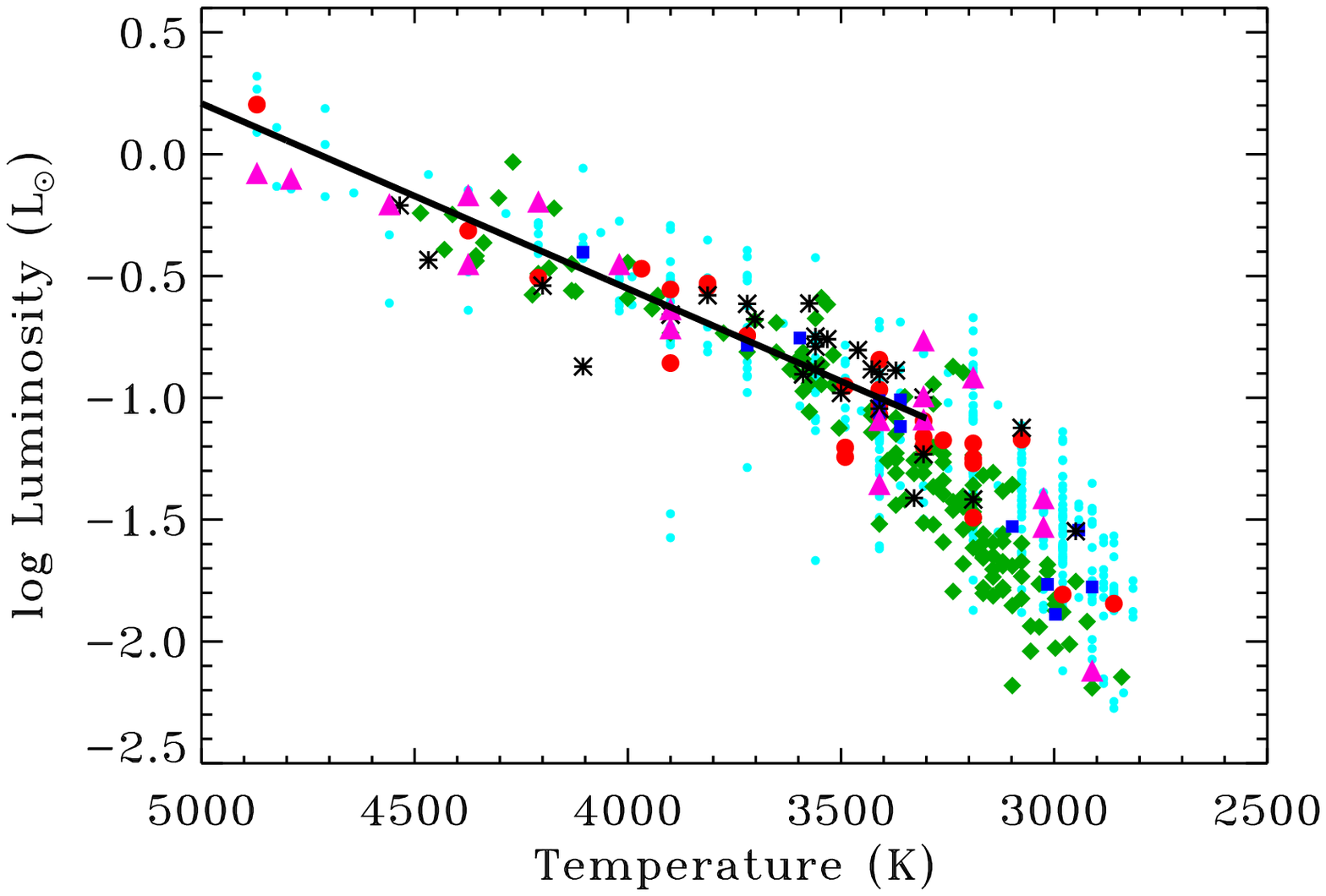}{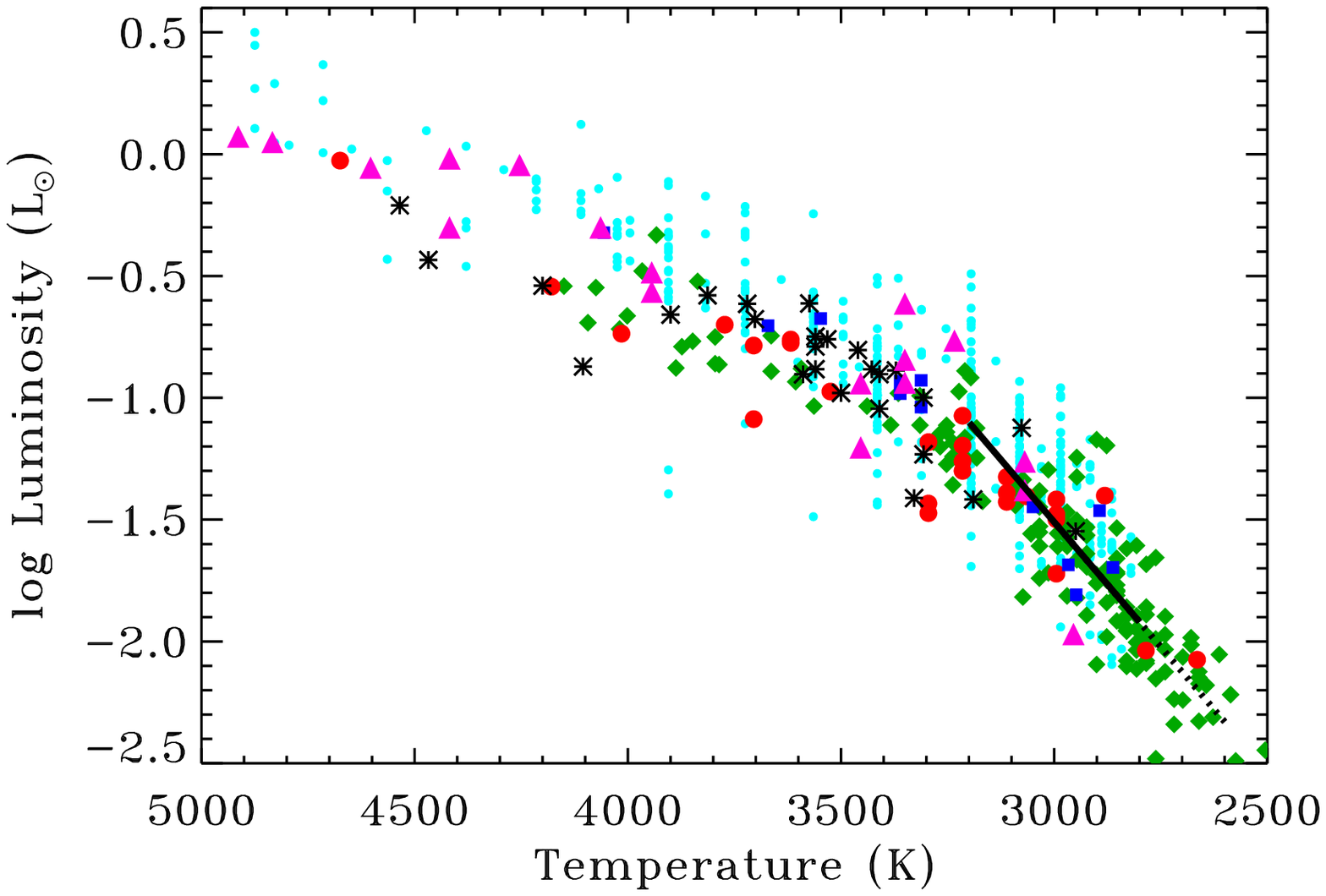}
\caption{HR diagrams for nearby young associations.  The top diagram
  shows the luminosity and temperature for members of the benchmark five
  associations (green diamonds for the Tuc Hor Moving Group, red circles
  for the $\beta$ Pic Moving Group, black asterisks for the TW Hya
  Association, blue squares for the $\eta$ Cha Association, purple
  triangles for the $\epsilon$ Cha Association, and small cyan circles
  for the Upper Sco OB Association).  The bottom panels
  show the same figure, with the stars offset in luminosity (left) to calculate a
  best fit to the 3200--5000 K stars (black solid line) and offset in temperature (right)
  to calculate a best fit to the 2800--3200 K brown dwarfs (black solid line, extends to $<2800$ K because some
  temperatures are shifted into this range).}
\label{fig:hrall}
\end{figure*}

Figure~\ref{fig:hrall} shows the combined HR diagrams for the five
benchmark young associations. 
The stellar locus between 3500--5000 K has a similar shape for all
clusters, with offsets in luminosity and temperature.
Stars in the older associations have systematically fainter luminosities
and warmer temperatures relative to young associations.
The slope of the locus is steeper for very low mass stars and
brown dwarfs$^7$ ($T<3200$ K for the young clusters and $<3400$ K for
older clusters) relative to the locus for solar mass stars.  The location where the break in slope of the
temperature-luminosity isochrone occurs is referred to here as the inflection
point in the isochrone.
\footnotetext[7]{These objects include very low mass stars and
  brown dwarfs and are termed brown dwarfs in this paper
  for ease of terminology.  Object masses are not calculated in this
  paper, and the distinction between stars and brown dwarfs is
 not relevant at young ages.}

The observed HR diagrams are consistent with the basic shape
predicted from isochrones obtained from pre-main sequence evolutionary
models (see comparisons in \S 4 and Fig.~\ref{fig:hrt2}). The gradual decrease in luminosity from 5000 to $\sim 3300$ K
is roughly consistent with all pre-main sequence evolutionary models. At
temperatures cooler than $\sim 3300$ K, the steep dropoff in luminosity
is also roughly consistent with predictions from
tracks that calculate pre-main sequence evolution for stars $<0.2$
$M_\odot$.   
Based on the HR diagrams, the inflection point where the slope
steepens 
occurs at ~$\sim 3400$ K for the older two associations and $\sim 3200$ K for younger
associations.  The temperature intervals are therefore different in
fits to stellar and brown dwarf loci of the older and
younger associations.

The
similarity in stellar loci suggests that the members may be analyzed
together after accounting for age-dependent differences between the
clusters.  In the following subsections, we
combine the loci from different associations into a single stellar
locus to calculate an empirical isochrone, which can then be
generically applied to any young cluster.  
Combining data from
different clusters yields results that are more robust than those calculated for
individual, sparsely populated associations.  When comparing cluster
ages, scatter in data points and different spectral
type distributions of members can introduce systematic errors in
relative ages.  

Once a generalized fit to temperature and luminosity is obtained,
scaling the slopes to each cluster in the luminosity direction 
then provide self-consistent comparisons
of relative ages that furthermore are robust to certain types of scatter in the temperature-luminosity
data points.  For example, linear fits to the TWA and Tuc-Hor
populations yield median luminosities of $\log L/L_\odot=-0.34$ and
$-0.70$, respectively, for 4200 K stars, or $-1.00$ and $-1.34$ for
3400 K stars$^8$.  These
luminosities can be compared among clusters to infer relative 
ages in $\log L$ as a fuction of $\log T$.
\footnotetext[8]{These luminosities differ from the final values listed in
  Table~\ref{tab:lumorder.tab} because these values are calculated
  from best-fit slopes for the two clusters, while the final values
  are obtained from scaling a line with a slope that best fits all
  regions together.}

The loci of stars above and below the inflection point (3200 K for
young regions, 3400 K for the older regions) are discussed
separately in \S 3.1 and \S 3.2.
 An underlying assumption to this approach is that the slopes of stellar loci are
constant longward and shortward of the inflection point and that the
slopes do not change with cluster age.  These sections include a
discussion of whether this simplifying assumption is correct.  For
brown dwarfs in particular, the $\log L-T$ slopes are different for
well-populated loci of Upper Sco and Tuc-Hor.  
The location of the inflection point ranges from 3000--3500 K and is discussed in
\S 3.3.

\begin{table}
\caption{Slopes of Linear Fits to Young Stellar Isochrones}
\label{tab:slopes.tab}
\begin{tabular}{lccccc} 
&&   \multicolumn{2}{c}{3200-5000 K}& \multicolumn{2}{c}{2800-3200 K} \\
\multicolumn{2}{l}{Empirical Fit} & Slope & StD & Slope & StD \\
\hline
\multicolumn{2}{l}{Combined} & $0.76\pm0.07$ & 0.15 & $2.05\pm0.15$ &  0.18 \\
\hline
\multicolumn{2}{l}{Young Groups Only$^a$}& $0.76\pm0.06$ & 0.16 &  $3.1\pm0.8$ & 0.20 \\
\multicolumn{2}{l}{Old Groups Only$^b$} & $0.85\pm0.08$ & 0.14 & $2.05\pm0.18$ & 0.18 \\
\hline
\multicolumn{2}{l}{Tuc Hor} & $0.83\pm0.12$ & 0.14 & $2.28\pm0.25$ & 0.18 \\
\multicolumn{2}{l}{BPMG} & $0.91\pm0.12$ & 0.14 & $1.87\pm0.38$ & 0.16\\
\multicolumn{2}{l}{TWA} &  $0.82\pm0.15$ & 0.17 & $^d$& \\
\multicolumn{2}{l}{$\epsilon$ Cha} & $0.73\pm0.08$ &0.17 & $^d$ & \\
\multicolumn{2}{l}{$\eta$ Cha} & $0.93\pm0.27$ & 0.06 & $^d$ & \\
\hline
\multicolumn{2}{l}{Upper Sco} & $0.85\pm0.07$ & 0.18 &  $3.5\pm0.8$
& 0.27\\
\hline
\hline
Model & Age$^c$ & 3400--5000 K &  & 2800--3200 K &  \\
\hline
BCAH98 & 3  & 1.04 &  & 1.97 & \\
BCAH98 & 20& 0.89 &  & 1.62  & \\
BCAH98-1.9  & 3   & 0.69 &  & -- & \\
BCAH98-1.9  &  20& 0.76  &  &  -- &\\
BHAC-2015 & 3 & 0.68 & & 1.49\\
BHAC-2015 & 20 & 0.72 & & 1.84\\
Pisa  & 3      &0.71 &  & -- & \\
Pisa  & 20  &0.76 &  & -- & \\
Dartmouth & 3 & 0.73 &   &  --  &\\
Dartmouth & 20   & 0.74 && -- &\\
Feiden & 3 & 0.75 & & 1.57 &\\
Feiden & 20 & 0.84 & & 2.02 & \\
Siess & 3  & 0.87 & & -- &\\
Siess & 20 & 1.00 & & -- &\\
DM97 & 3    & 0.70   &  & 2.46  & \\
DM97 & 20  & 0.85   &  & 2.94 &  \\
PARSEC & 3 & 0.62  &    & 1.15 & \\
PARSEC & 20 & 0.51 &   & 1.32 & \\
PARSECv1.1 & 3 & 0.83   &   & --& \\
PARSECv1.1& 20 & 1.09 &    & & --\\
\hline
\multicolumn{6}{l}{Slope (in $\log L$ per 10$^{3}$ K), uncertainty,
  and standard deviation}\\
\multicolumn{6}{l}{of fits to individual, combined, and model isochrones}\\
\multicolumn{6}{l}{$^a$$\eta$ Cha, $\epsilon$ Cha, and TW Hya
  Associations}\\
\multicolumn{6}{l}{$^b$$\beta$ Pic and Tuc Hor Moving Groups}\\
\multicolumn{6}{l}{$^c$Age of the model isochrone in Myr}\\
\multicolumn{6}{l}{$^d$Insufficient number of data points}\\
\end{tabular}
\end{table}

\begin{figure*}[!th]
\epsscale{1.}
\plotone{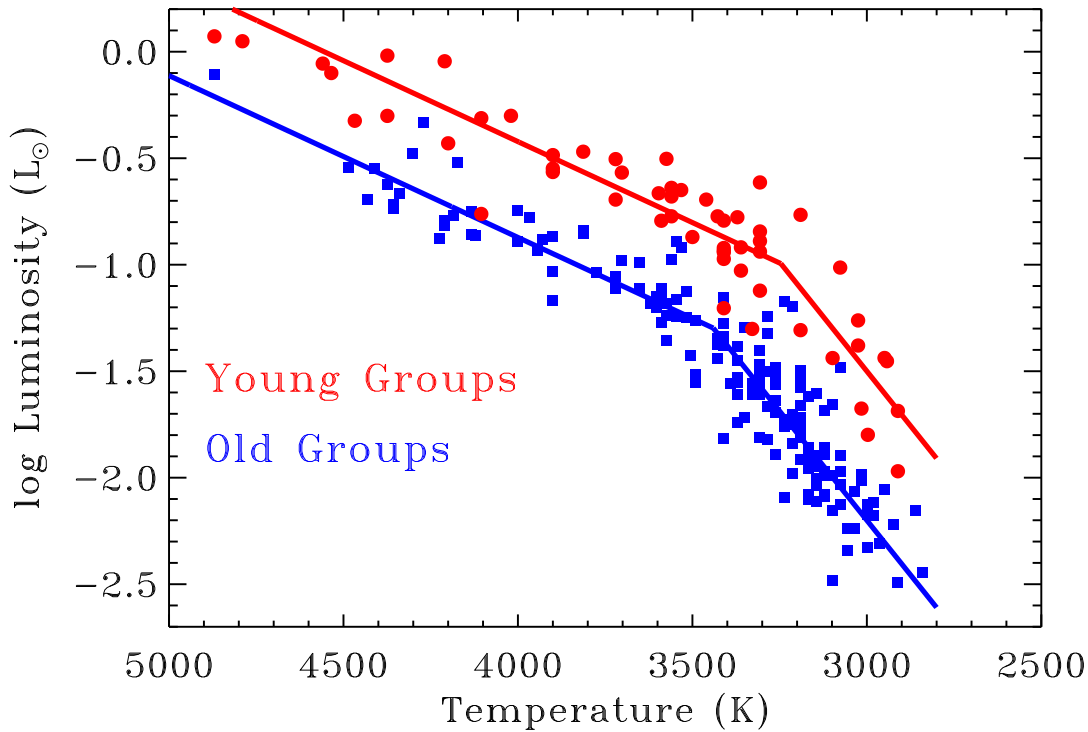}
\caption{The HR diagram for young (red circles) and old clusters (blue
  squares), with members of the $\beta$ Pic
  Moving Group shifted to overlap with the Tuc Hor stellar locus,
  while $\eta$ Cha and TW Hya Association members shifted
  to overlap with the $\epsilon$ Cha Association stellar locus.  The blue and red lines show
  the best fits to the combined low-mass star and brown dwarf locus,
  also shifted to match the young and old groups.}
\label{fig:hrt}
\end{figure*}

\subsection{Empirical Isochrones for low mass stars}

In this subsection, we describe global fits to the 3200--5000 K 
region (3400--5000 K for the $\beta$ Pic and Tuc-Hor Moving Groups), which corresponds to M3--K0 spectral types.
Since the empirical loci in this temperature range are similar, a global linear fit to the
entire dataset provides a consistent assessment of the relative luminosity
evolution of a given cluster.  
In fits to the combined HR diagram of
the five benchmark associations, the free parameters are the slope of
the line and a different y-intercept for each cluster.  For
illustrative purposes, the different y-intercepts are described here
as luminosity offsets.

The lower left panel of Figure~\ref{fig:hrall} shows the HR diagram,
with each association shifted by the respective
difference in y-intercept to match the location of the TWA.  The empirical isochrone for the combined HR diagram of nearby
associations has a best fit slope
of 
\begin{equation}
\log L/L_\odot\propto (0.76\pm0.07)~{\rm T}/10^3~{\rm K~(T_{{\rm infl}}-5000~K)},
\end{equation}
based on linear regressions that includes errors in $T$ and $L$ (as
described in \S 2.3) and calculated using 
mpfitexy.pro in IDL \citep{Williams2010}.    The inflection point
$T_{{\rm infl}}$ used for the fits here is 3200 K for the younger groups and 3400 K
for the older groups.
The 0.07 dex uncertainty
includes a 0.05 dex statistical uncertainty and a $\sim 0.05$ dex
uncertainty based on different linear fitting techniques, how the data
points are weighted, and the inclusion or
exclusion of deviant data points.  This uncertainty
does not include possible errors in the spectral type-temperature
conversion or in bolometric corrections, which may depend on spectral
type.  
The standard deviation in luminosity between
the observations and best linear fits is $\sim 0.15$ dex.

The same analysis calculated from the \citet{Pecaut2013} spectral type-temperature conversion and
bolometric corrections produces a slope of $0.72\pm0.04$ $\log L$ per 10$^{3}$ K, which is formally consistent with the results when
using the \citet{Herczeg2014} conversions.
The different slopes
would yield a change in $\log L$ of $0.07$ dex between the hottest and coolest
stars in the applicable temperature range, which is smaller
than other systematic errors.

Table~\ref{tab:slopes.tab} lists the best fit slopes for linear fits
to the the combined data set, to individual clusters, and to the
clusters combined into the young and old groups.
When shifted based on the best fit offsets, the three
youngest benchmark associations have steeper slopes than the older
clusters, which suggests some evolution with age.  The difference in
slopes between the young and old clusters, 0.76 and 0.85 $\log L$ per 10$^{3}$ K,
may lead to systematic differences as large as 0.14 dex for
stars at 3400 K compared with 5000 K.

\subsection{Empirical Isochrones for Young Brown Dwarfs}

In this subsection, we describe global fits to objects cooler than the
inflection point in the isochrones (2800--3400 K for the Tuc-Hor and
$\beta$ Pic Moving Groups, 2800--3200 K for the TW Hya, $\eta$ Cha, and
$\epsilon$ Cha Associations).
For illustrative purposes, the different
y-intercepts are visualized as temperature offsets so that the loci
overlap (lower right panel of Fig.~\ref{fig:hrall}).  This
visualization is practical because the inflection point occurs at
different temperatures but approximately the same luminosity (to
within a few tenths of dex; see also similar behavior in the model
isochrones in Figs.~\ref{fig:baraffefeiden} and \ref{fig:hrt2}).

The
total combined fit to our benchmark clusters yields a slope of 
\begin{equation}
\log L/L_\odot \propto (2.05\pm0.15)~{\rm T}/10^3~{\rm
  K ~~(\rm T_{{\rm infl}}}-5000~K),
\end{equation}
where the temperature range goes from 2800--3200 K for the younger clusters and 2800--3400
K for the older clusters (as also described above with $T_{{\rm infl}}$).  The error in the
slope is calculated as in \S 3.1.    The fits are dominated by members of the Tuc Hor and $\beta$
Pic Moving Groups because those regions have more identified brown
dwarf members.  Fits are calculated for the combined locus of young
associations but are not calculated for each individually because each
association  has only a few data points.
Table~\ref{tab:slopes.tab} lists slopes for the individual regions and
the combined HR diagram.  

The coolest known objects in these regions ($<2800$ K) are fainter than
expected from these linear fits.  They are excluded in this analysis,
but their inclusion would significantly steepen the slopes of this
lines.  This discrepancy suggests that these fits may not be
robust to the selection of the low temperature cutoff for this fit.

The measured uncertainty in the linear fit of 0.15 $\log L/L_\odot$
per $10^3$ K leads to a 0.06 dex uncertainty in relative luminosity
between a 3200 K and 2800 K brown dwarf.  The difference in
slopes between the older and younger associations likely leads to 0.09
dex uncertainties over the same regions.  This difference is likely
real evolution in the brown dwarf loci and not related to uncertainty in the fits.   In \S 3.4, comparisons to the Upper Sco OB
Association demonstrate that the slope becomes shallower with age.
The results from linear fits in producing a luminosity for a 3000 K
brown dwarf may still be robust as long as the
known membership is well distributed between 2800--3200 K.  However,
this relationship should only be used with caution.

The fit to the brown dwarf locus is not recalculated for
\citet{Pecaut2013} parameters because their bolometric corrections are
not calculated for spectral types later than M5.

\subsection{Combined Isochrones between 2800--5000 K}

Subsections \S 3.1 and 3.2 describe empirical fits to HR diagrams,
as illustrated in Figure~\ref{fig:hrall}.  The regions warmer and
cooler than the inflection point in the temperature-luminosity locus
were fit separately, with one slope but different y-intercepts
for each association.   The shifts obtained from the y-intercepts 
are visualized as luminosity shifts, while the fits to the
cooler stars are visualized as temperature shifts.

The location of the inflection point, which
motivates the separate fits to the two temperature ranges, depends
on the cluster age.  In this subsection, we compare the independent
results from the fits to low mass star and brown dwarf isochrones to
describe how the empirical lines combine to describe the low mass
locus of young clusters.  
For illustrative purposes only, the combined isochrones are plotted
against the young and old associations in Figure~\ref{fig:hrt}.

\begin{figure}[!t]
\epsscale{1.}
\plotone{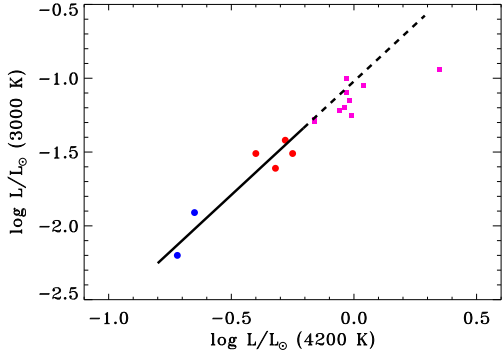}
\caption{The relationship between the luminosity for 4200 K and 3000 K
stars based on the lines fit to the individual
clusters.  The blue and red circles are the old and young associations
(including Upper Sco), respectively, and are best fit with the solid
black line.  The purple squares are preliminary results from younger
clusters, most of which are fit by an extrapolation of the best fit
line to higher luminosities.}
\label{fig:shift}
\end{figure}

Figure~\ref{fig:shift} shows the luminosity of 3000 K brown dwarfs
versus 4200 K stars from our independent fits to the young clusters
with a best fit line of
\begin{equation}
\log (L_{4200}/L_\odot)= -1.02 +1.54 \log(L_{3000}/L_\odot).
\end{equation}
{\it Combining the line of equation 3 with the two slopes reported in
equations 1 and 2 fully describes the evolution of the empirical locus of single
low mass stars and brown dwarfs
in clusters.}  The isochrone for low mass stars falls in luminosity
more quickly than that for brown dwarfs.  The inflection point where
the  temperature-luminosity slope changes is determined by the
intersection of the stellar and brown dwarf locus described by this line.  

As a further test of our method, Figure~\ref{fig:shift} also shows preliminary
results of its application to younger clusters with proportionally
higher luminosities.  The data are obtained for IC 348, Chameleon I,
Lupus I, Taurus, $\sigma$ Ori, $\lambda$ Ori, and the ONC \citep{Luhman2003,Luhman2007,Luhman2010,Rebull2010,Rigliaco2011,Bayo2011,DaRio2012}.
Although the errors in placement of individual
stars on the HR diagram are larger for younger stars with more
influence from disks and heavier extinction, their mean luminosities 
are encouragement that our method could be applied to younger regions in order to constrain the
luminosity evolution for comparison to theoretical models.

\subsection{Application to the Upper Sco OB Association}

\begin{figure*}[!th]
\epsscale{1.}
\plottwo{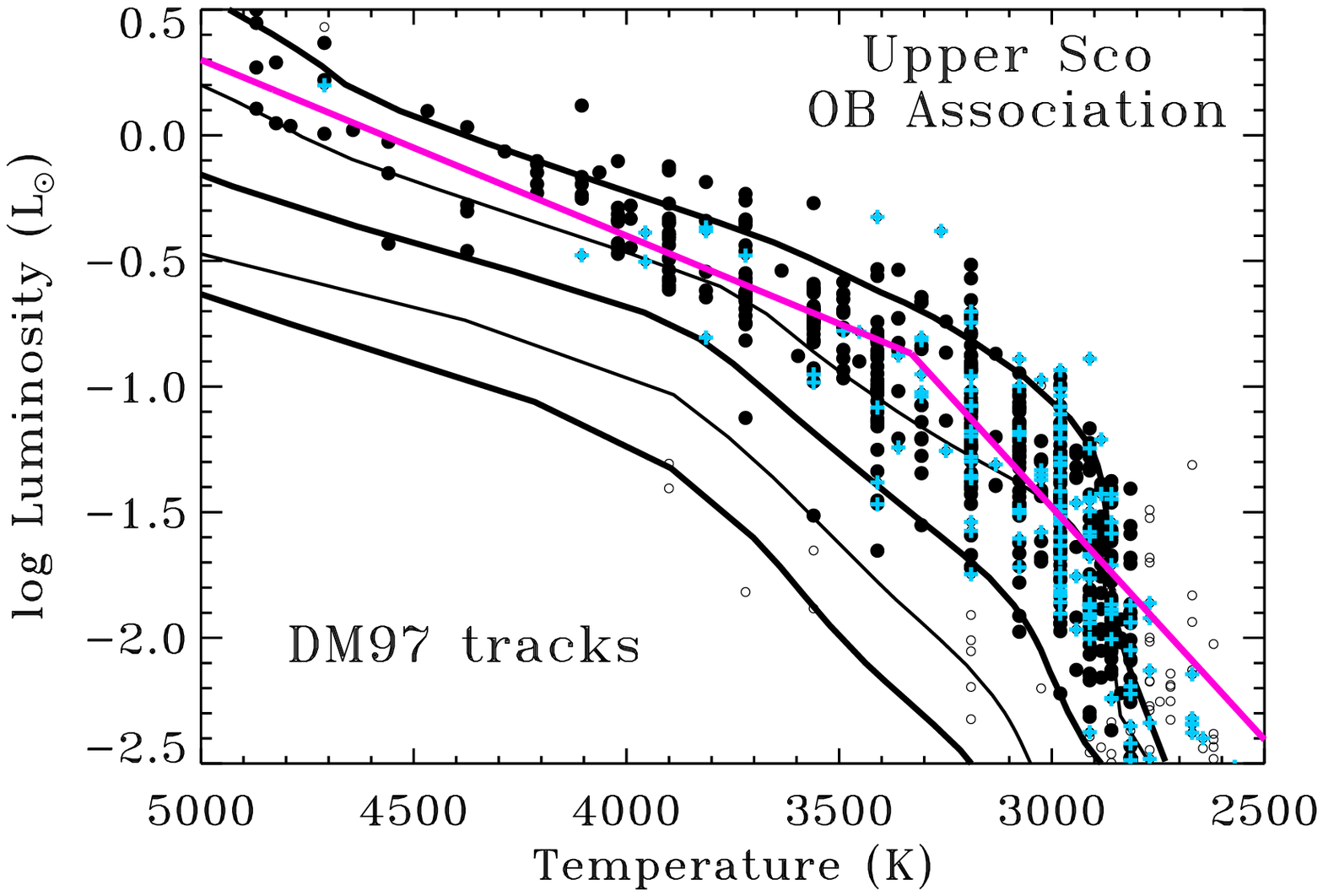}{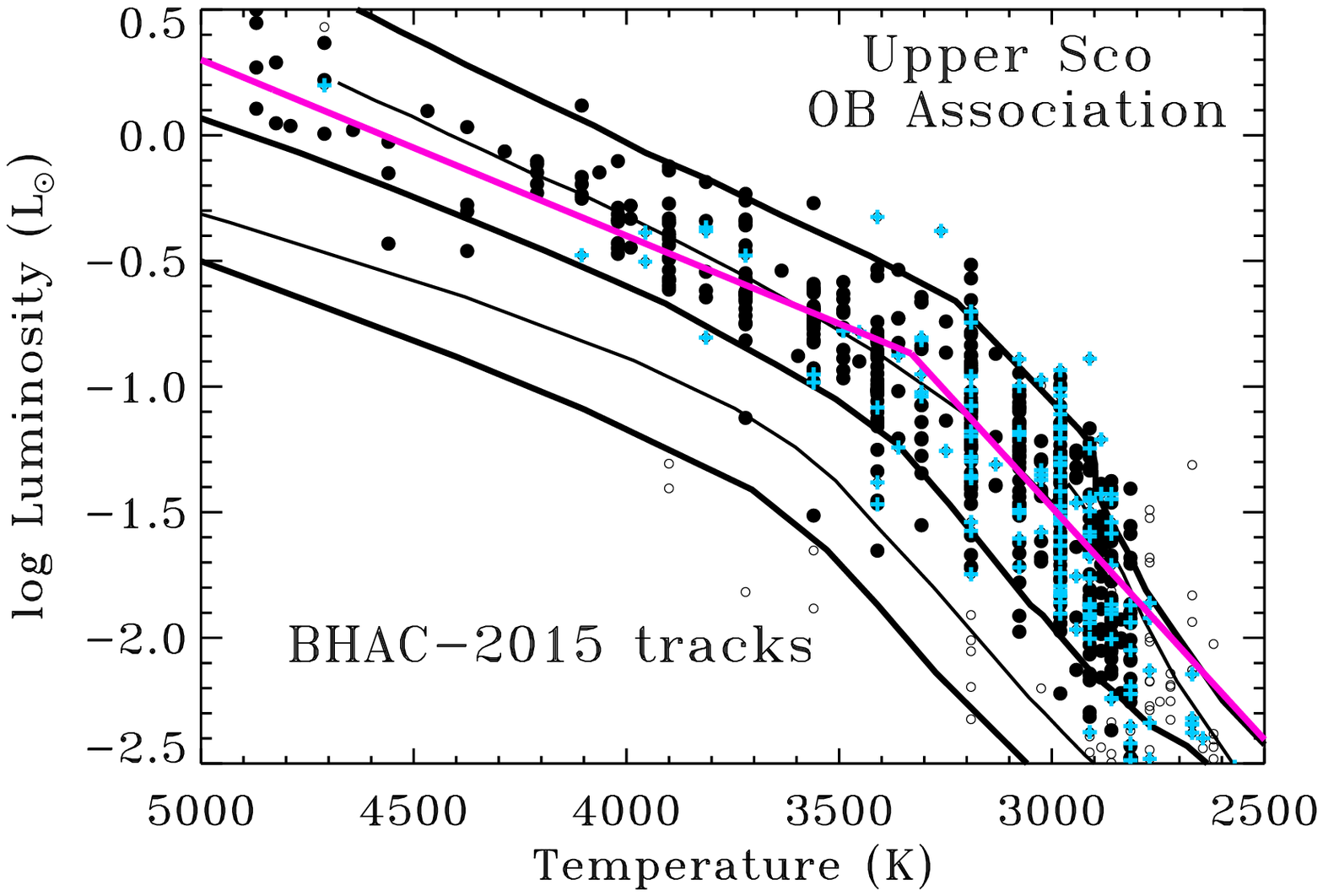}
\caption{The HR diagram of the Upper Sco OB Association, with
  empirical linear fits scaled in two temperature ranges to the stellar locus.  Data points used
  for fits are solid circles.  Objects with that are cooler than 2800
  K or are more than 0.8 dex fainter than the median are not fit
  (empty circles).  Objects with disks (cyan plus signs) are also
  excluded in the fits.  The black lines show
isochrones of $\log$ age = 6.0, 6.5, 7.0, 7.5, and 8.0, with 6.0, 7.0,
and 8.0 in bold.  The purple line shows the empirical isochrone for
low mass stars and brown dwarfs, scaled to the Upper Sco OB
Association data.}
\label{fig:usco}
\end{figure*}

Figure~\ref{fig:usco} shows how the fits derived above are applied
to the HR diagram of the Upper Sco OB Association.  Lines with slopes
that best fit the loci of low mass stars (3200--5000 K) and brown dwarf/very low
mass stars (2800--3200 K) are scaled to the measured
isochrone.

The initial sample includes stars with K-W3 color consistent with
diskless stars and also stars without WISE detections.  An initial scaling to the
3200--5000 K region (186 stars) has $\log
L_{4200}/L_\odot=-0.28$ with a standard deviation of 0.26 dex, much larger than that obtained for the other
clusters.  Several stars are severe outliers, defined here as 0.5 dex
from the $\log L$ calculated in the initial fit.  Our adopted fit of
176 stars excludes these severe outliers.
We calculate $\log L_{4200}/L_\odot=-0.23$ with a
luminosity scatter of 0.18 dex.  Fitting stars from 3400--5000 K
yields the same $\log L_{4200}$.
This luminosity spread of 0.18 dex is slightly larger than that calculated for
the other associations, with extra scatter that may be caused by a real age
spread in the cluster or, as suggested by \citet{Slesnick2008}, a
$\sim 30$ pc depth of the region.

The brown dwarf locus of Upper Sco at 2800--3200 K has a much steeper slope than
the linear fit to the 5--40 Myr old sample, which points to problems
in our simplifying assumption that the slope of the 2800--3200 K
object locus does not evolve on these timescales. 
The fit to the slope of combined loci of the five benchmark clusters is dominated by Tuc-Hor members.
 However, the slopes from
the younger associations are steeper than for the older associations.
These differences suggest that the empirical brown dwarf locus is better
described by a swinging gate rather than a single slope.  This same
swinging gate behavior is also
suggested by the theoretical isochrones covering this low temperature
regime (Figure 9), where the predicted slopes appear to flatten with
increasing age.  In particular, the brown dwarf loci is
remarkably well described by the DM97 isochrones and is also consistent with the BHAC-15 isochrones.

While a
more in-depth analysis is beyond the scope of the simple-minded spirit
of this paper, these problems with the brown dwarf isochrones lead us
to focus subsequent analysis of the correspondence between stellar mean stellar luminosities and stellar ages on the 3200--5000 K
stellar locus.

\subsection{Ordering Clusters by Luminosity}

In this section, we apply the analytical approximations of isochrones
from \S 3.1--3.2
to the set of young clusters to order them in luminosity, as a proxy for age.
The fits to luminosities for 3200--5000 K stars (\S 3.1) are likely more
accurate than for the 2800--3200 K brown dwarfs (\S 3.2) because
an analysis of the brown dwarf locus relies heavily on the two older
moving groups and may not accurately describe the brown dwarf locus of
younger clusters.

\begin{table}[!t]
\caption{Luminosity Ordering of Young Clusters}
\label{tab:lumorder.tab}
\begin{tabular}{lc|ccc|ccc} 
Stellar & & \multicolumn{3}{c}{Low mass stars$^a$} & \multicolumn{3}{c}{Brown
   Dwarfs$^b$} \\
Ass'n & d$^c$ & N & $\log L_{4200}$ & StD &
N& $\log L_{3000}$ & StD \\
\hline
Upper Sco     & 145 & 176 & -0.25 & 0.18 & 286 & -1.51 & 0.27\\
$\epsilon$ Cha & 111 & 14 & -0.28  &0.17 & 4 & -1.42 & 0.23 \\
$\eta$ Cha  & 94 &  7 & -0.32  & 0.06 & 5 & -1.61 & 0.16 \\
TW Hya      & (57) & 22 & -0.40 & 0.17 & 3 & -1.51 & 0.22 \\
$\beta$ Pic & (30) & 16 & -0.65 & 0.14  & 11 & -1.91 & 0.15 \\
Tuc Hor      & (51) & 48 & -0.72 & 0.14 & 89 & -2.20 & 0.18\\
\hline
\hline
\multicolumn{8}{l}{$^a$From linear fits to 3200/3400--5000 K
  stars, see \S 3.1}\\
\multicolumn{8}{l}{$^b$From temperture shifts for stars of 2800--3200
  K, see \S 3.2}\\
\multicolumn{8}{l}{$^c$Distances in () are median association
distances}\\
%TW Hya:  -0.39 \pm 0.12 without TWA 9A/B
\end{tabular}
\vspace{10mm}
\end{table}

Table~\ref{tab:lumorder.tab} shows the ordering of clusters by
luminosity for the 3200--5000 K fits.  
The five benchmark associations are ordered in luminosity as
expected from previous studies (e.g., see discussions in Mamajek 2008;
Hillenbrand 2009; Kraus et al.~2011
and Fang et al.~2013).  The Upper Sco, $\epsilon$ Cha and $\eta$ Cha Associations are
the youngest, followed by the TW Hya Association, the $\beta$ Pic
Moving Group, and the Tuc-Hor Moving Group.  Use of \citet{Pecaut2013}
temperatures and bolometric corrections would increase association
$\log L_{4200}$ by 0.02-0.03 dex.

The formal uncertainties in the luminosity (age) ordering are calculated from a combination
of (1) the standard
deviation divided by the square root of the number of stars in the
sample, and (2) systematic uncertainty in distances. 
Methodological differences in luminosity calculations and in the
census of multiplicity also can
affect the luminosity ordering of clusters.  For the specific
benchmark clusters analyzed here, the age ordering is secure, with the
possible exception of the $\eta$ Cha Association.  To swap the TWA and
$\eta$ Cha Associations would require a systematic change in
$L_{4200}$ by 0.08 dex (20\%).  For young clusters, the primary
  culprits to explain such a luminosity change of 0.1 dex would be systematic errors in the distance (12\% change),
  extinction (change of $A_V=0.9$ mag), and spectral type (change of
  140 K).  \citet{Weinberger2013} argue that TWA 9A and 9B should be
  excluded from the TWA census because they are much fainter than
  expected, which would increase the cluster luminosity by 0.04 dex.
These comparisons are also affected by differences in
binary accounting and by accretion, if samples include disk sources.

\begin{figure}[!t]
\epsscale{1.1}
\plotone {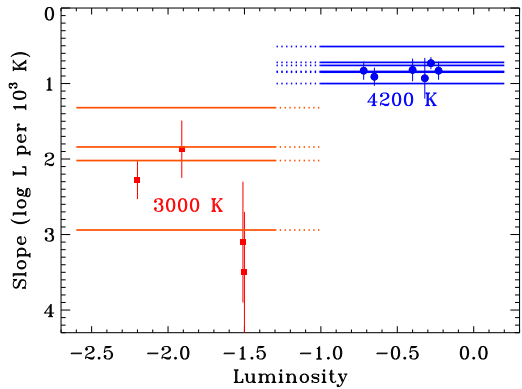}
\caption{The measured slopes of luminosity versus temperature for each association (blue circles for 3200--5000
K stars with luminosities assessed at 4200 K; red circles for brown
dwarfs with luminosities assessed at 3000 K) compared with slopes predicted
from Feiden, BHAC-15, Pisa, DM97, PARSEC, and Siess tracks
evaluated at 20 Myr (see Tables 3--4 for values).
The solid lines show the luminosity interval over which the slopes
apply for both the young and old associations in this paper.  The
dotted lines show the intermediate regime where the inflection point
occurs.  The brown dwarf points include one slope and luminosity calculated for the
combined young associations.}
\label{fig:lumslope}
\end{figure}

\section{Comparing Empirical and Model Isochrones}

The relationship between stellar luminosity and
effective temperature has been measured for six young associations.
Because these stars
are contracting pre-main sequence stars, the radius and therefore
luminosity decreases with time.  The luminosity of stars at a fixed
temperature is therefore used in \S 3.5 as a proxy for age. 
The
conversion from luminosity to age depends on the adopted choice among the many
evolutionary models, as shown in Fig.~3, and on the stellar temperature.
At a fixed temperature, the age
estimates differ between the many evolutionary models.  Mismatches
between empirical and model isochronal slopes lead to different age estimates along an
empirical isochrone when a
given model family is applied to stars with different temperatures.

\begin{figure*}[!t]
\epsscale{1.1}
\plottwo{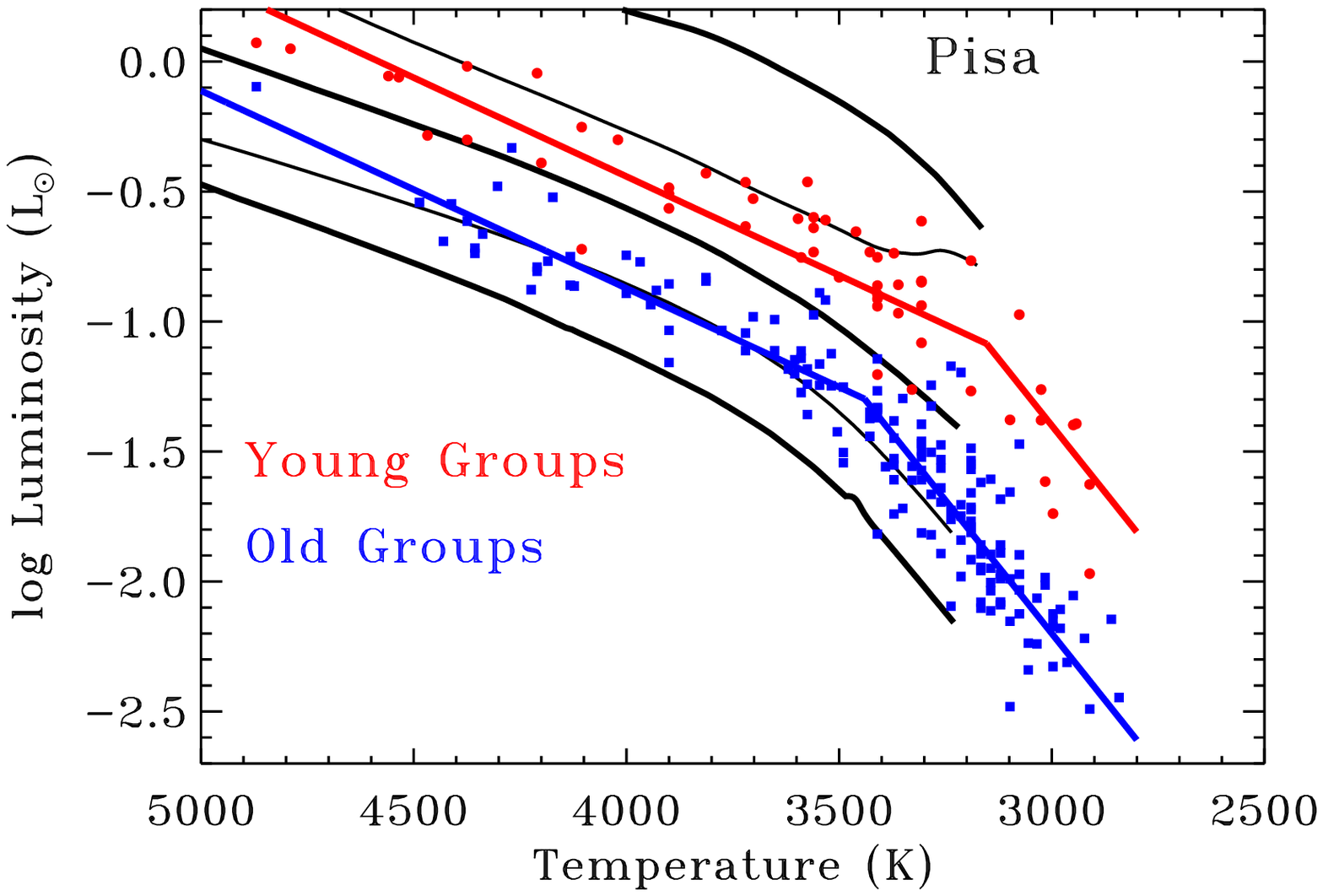}{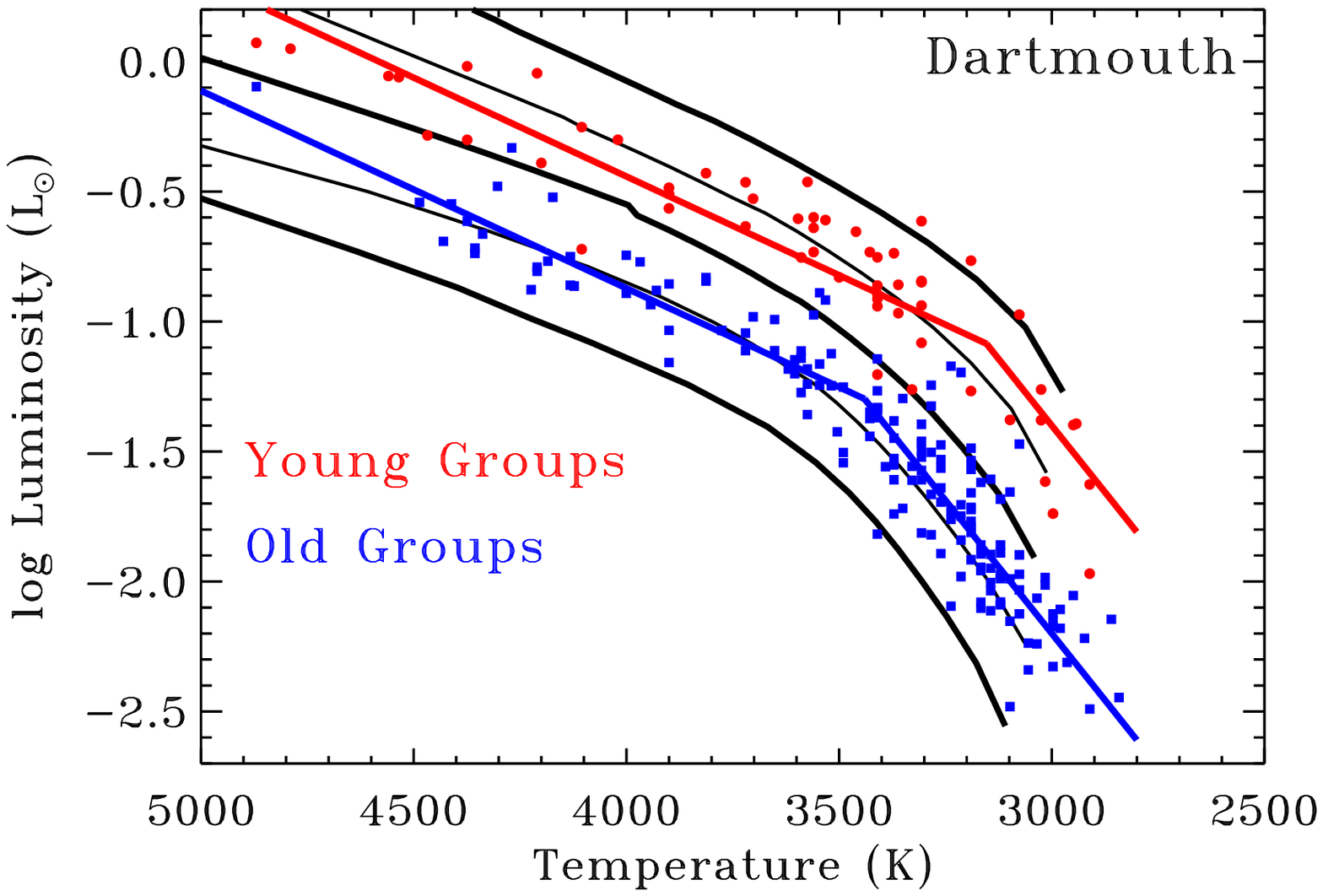}
\plottwo{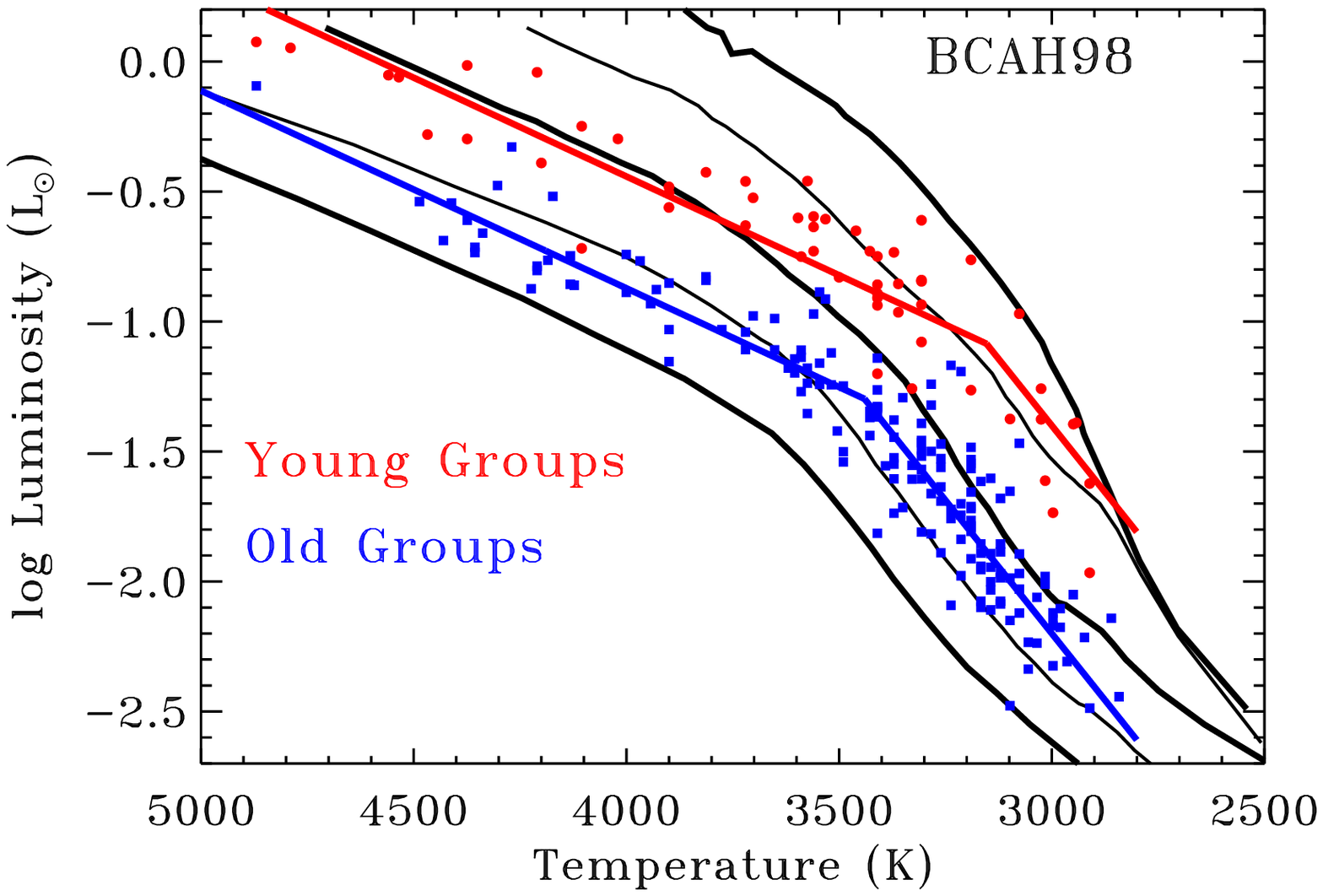}{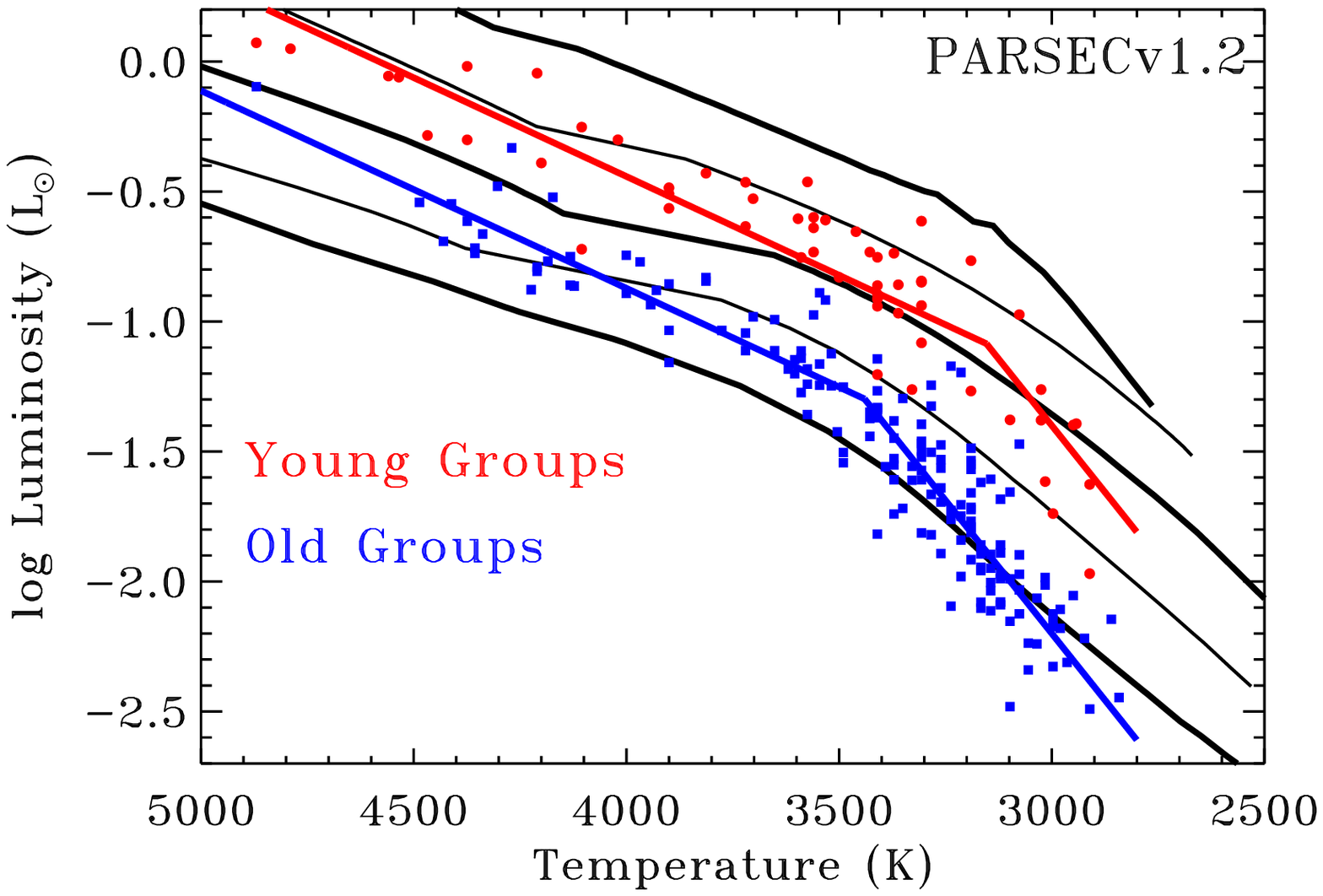}
\plottwo{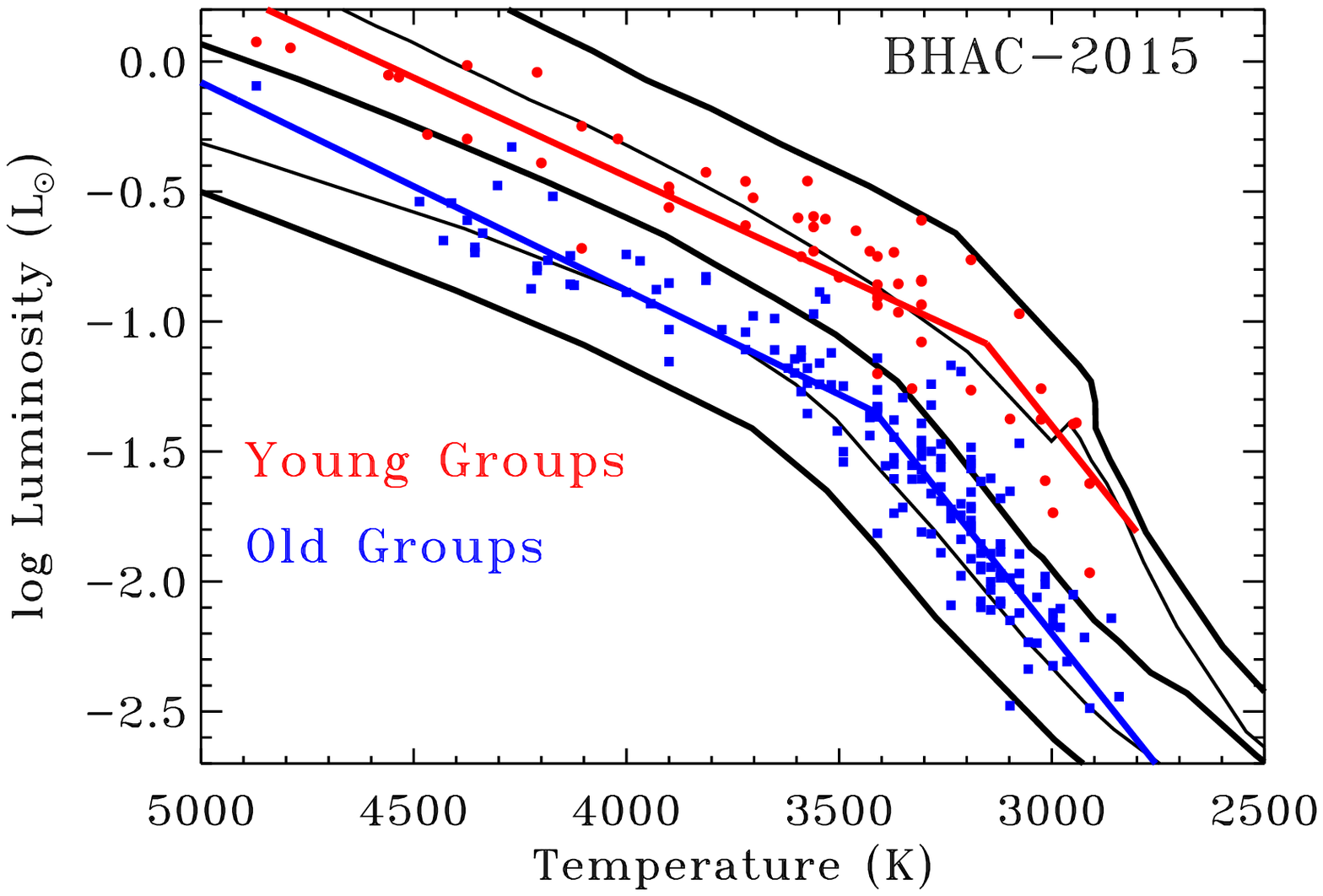}{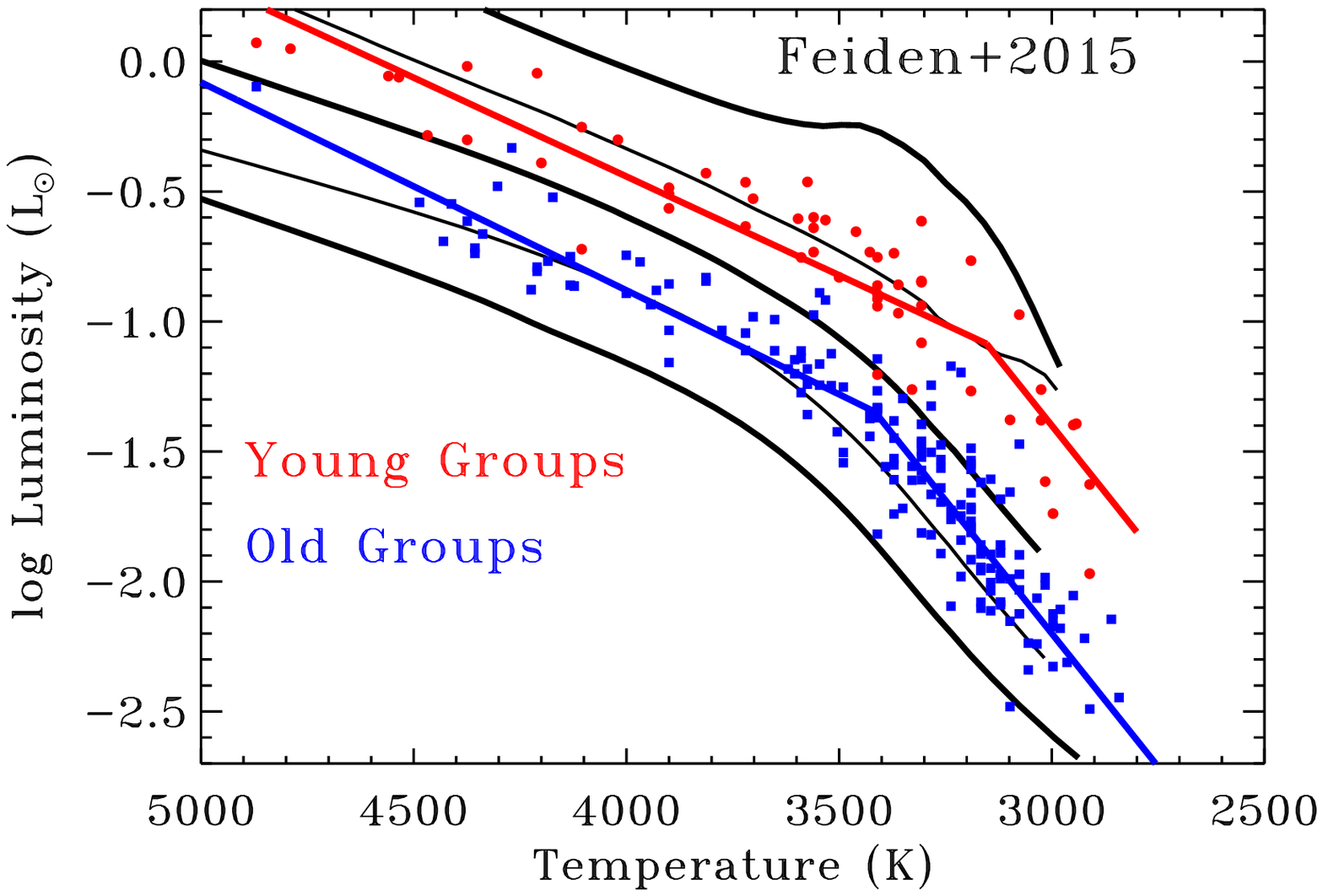}
\plottwo{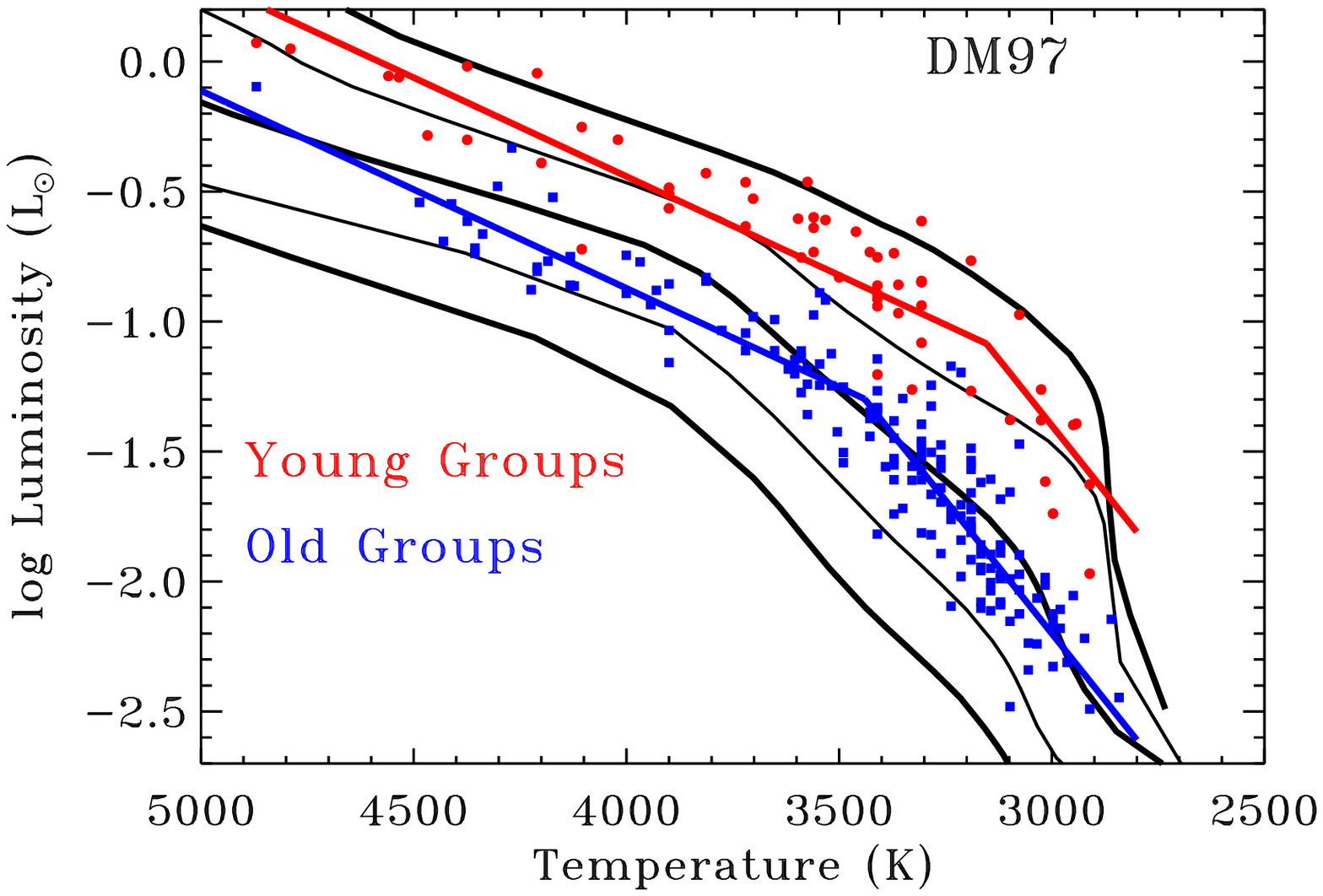}{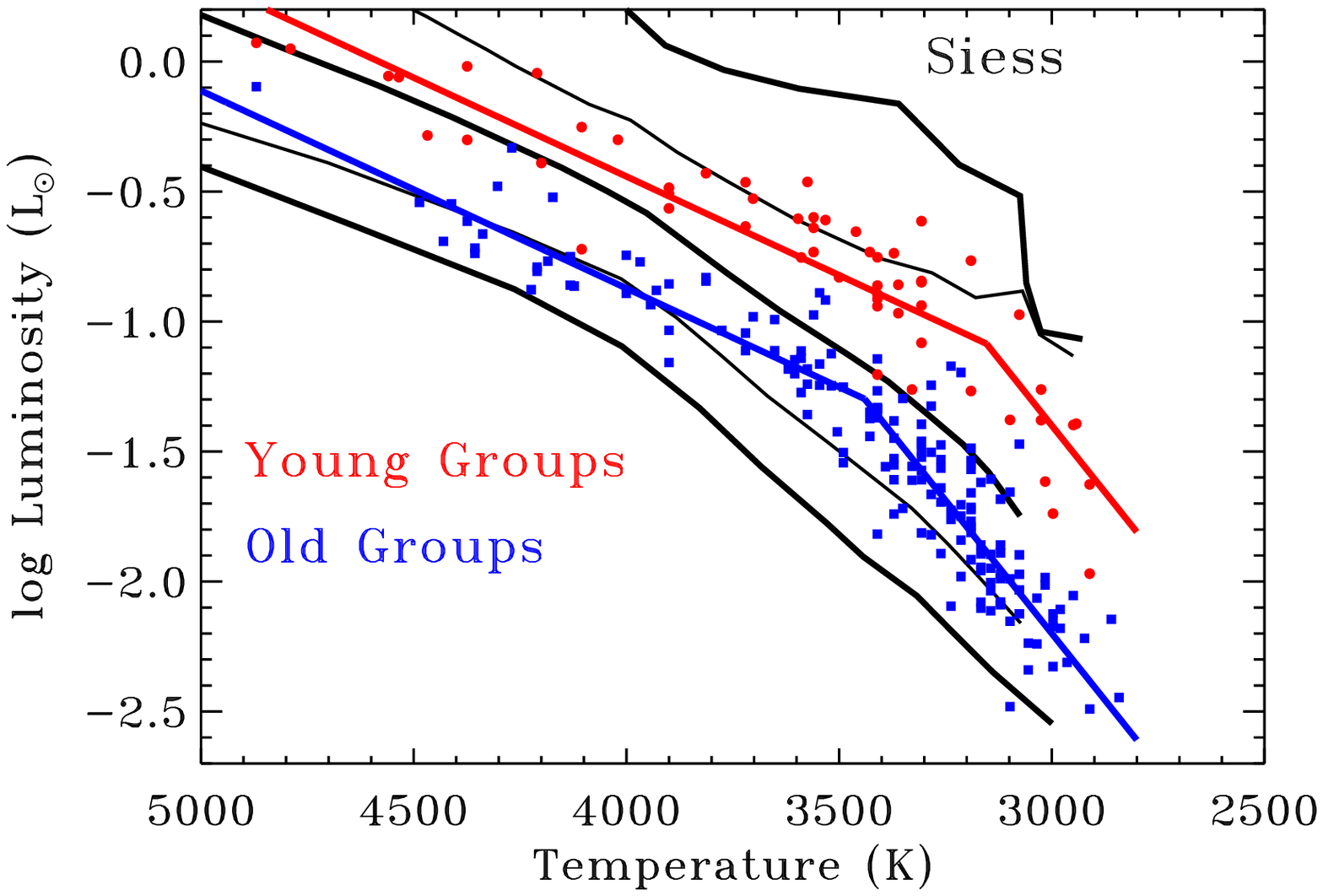}
\caption{The empirical HR diagrams, along with two-temperature regime
  linear fits,  compared to isochrones from
  pre-main sequence evolution models.  The points for $\eta$ Cha and $\epsilon$ Cha
  Association members are shifted to be consistent with the isochrone
  for TW Hya, while the $\beta$ Pic Moving Group members are shifted to
be consistent with Tuc-Hor Moving Group members.  As in Figure 8, the black lines show isochrones of log age
= 6.0, 6.5, 7.0, 7.5, and 8.0, with 6.0, 7.0, and 8.0 in bold.}
\label{fig:hrt2}
\vspace{5mm}
\end{figure*}

Figures~\ref{fig:lumslope}-\ref{fig:hrt2} and Table~\ref{tab:slopes.tab} compare the empirical isochrones to the
pre-main sequence models discussed in this paper.  The model
isochrones are curved but can be well approximated with a two
temperature fit within the 2800--5000 K region.  The slopes presented
in Table~\ref{tab:slopes.tab} are intended as a guide, with a measured
slope that is sensitive to the temperature region defined in the fit.
For low mass stars (3200--5000 K), the luminosity slopes in the
Dartmouth, Pisa, BCAH98-1.9, BHAC-2015, and Feiden
pre-main sequence tracks are consistent with the observations.  These
models produce similar ages for stars at 3800 and 4200 K.  On the
other hand, the isochronal slopes predicted from
the DM97, Siess, BCAH98, and PARSEC models differ from empirical slopes.
This discrepancy leads to differences in expected luminosities versus
spectral type.  The DM97, Siess, and BCAH98 models all have a
steeper slope than is observed, which leads to a temperature (mass)
dependence in age estimates.  As an example, if we normalize our
empirical isochrone and the BCAH98 isochrone to a young 4200 K star,
a co-eval star of 3400 K would appear $0.3$ dex more luminous  ($0.4$ dex
younger in age) than expected from the BCAH98 isochrone.
The PARSEC isochrones conversely have a shallower slope than is observed, which
leads to 3400 K stars appearing older than 4200 K stars.

The new BHAC-2015 and Feiden models have
isochrones that are 
similar to the empirical isochrones.  The inflection point in both
BHAC-2015 and Feiden occurs at temperatures $\sim 50-150$ K hotter than in the empirical
isochrones, which leads to mismatches between empirical and model
isochrones for very low mass stars and brown
dwarfs.  These two models predict isochrones that are close enough to
empirical isochrones that the differences could be explained if
effective temperatures below 3500 K are systematically underestimated
by 50-100 K.

\begin{table*}[!t]
\begin{center}
\caption{Consistent Sets of Ages (in Myr) of Young Associations$^a$}
\label{tab:ages.tab}
\begin{tabular}{lcc|ccc|cc|c}
 Stellar &  & &DM97 &  Pisa & BCAH98 &  Feiden &
 BHAC-2015 & Li Depl.$^b$ \\
Association & d (pc) &$\log L_{4200}/L_\odot$  &   3400 K & 4200 K &
4200 K & 3800 K & 3800 K & Age \\
\hline
    Upper Sco &  $145\pm10^c$ & $-0.25\pm0.06$ & $1.7\pm0.2$  &
    $5.0\pm1.1$ & $10.3\pm1.9$ & $4.2\pm1.0$ &  $4.1\pm1.0$ & --\\
 $\epsilon$ Cha & $111\pm6^c$&  $-0.28\pm0.06$  & $1.9\pm0.3$ &
$5.6\pm1.4$ & $11.3\pm2.1$  &  $4.7\pm1.2$ & $4.7\pm1.2$  & --\\
    $\eta$ Cha & $94\pm5^c$  & $-0.32\pm0.05$ & $2.1\pm0.3$ &  $6.6\pm1.3$ & $12.7\pm2.1$ & $5.5\pm1.3$
    &  $5.5\pm1.3$ & -- \\
          TWA &   (57) & $-0.40\pm0.04$ &    $2.7\pm0.4$  &
          $9.0\pm1.5$ & $16.5\pm2.1$   &     $7.9\pm1.4$ &
          $8.1\pm1.4$ & -- \\
        BPMG &  (30) & $-0.65\pm0.04$ & $6.1\pm0.7$ & $24\pm4$  &
        $36\pm5$& $23\pm4$ & $22\pm4$  &  $24\pm4$ \\
    Tuc Hor & (51)  & $-0.72\pm0.02$ & $7.7\pm0.5$  & $32\pm3$ &
    $44\pm3$ & $30\pm3$ & $29\pm3$ & $40\pm4$ \\
\hline
\multicolumn{9}{l}{$^a$For models that produce a
  younger, intermediate, and old age, plus the recent BHAC-2015 and
  Feiden models.}\\
  \multicolumn{9}{l}{~~~Age assessed at the listed temperature because of temperature
  dependence in age estimates.}\\
\multicolumn{9}{l}{$^b$Li depletion boundary ages, see text for details.}\\
\multicolumn{9}{l}{$^c$Estimated uncertainty in median cluster
  distances.}
\end{tabular}
\vspace{10mm}
\end{center}
\end{table*}

Given these mismatches between model and observed isochrones,
  even for the latest models for cool temperatures,
comparisons between clusters should be made at a constant temperature.
Age estimates for each model are evaluated at 3400, 3800, and 4200 K
and are plotted in Fig.~\ref{fig:ages_hist} for three of the
benchmark associations and for the Upper Sco OB Association.  
Because of the discrepancies in slope described above, for most models
the 3400 K stars are 0.2--0.5 dex younger than 3800 K and 4200 K
stars.  The PARSEC models have the reverse dependence, with the 3400 K
stars older than the 4200 K stars.

The brown dwarf ages are not plotted.  For the now-outdated BCAH98 models, ages of
3000 K stars are another 0.3 dex younger than the 3400 K stars.  On
the other hand, for PARSEC models ages of 3000 K stars are typically
0.1-0.3 dex older than 3400 K stars.  However, the differences between the brown dwarf loci of Upper Sco OB
Association and the Tuc Hor Moving Group demonstrates that the shape
and slope of the brown dwarf isochrone evolves with time.  Fitting a
single slope to the brown dwarf isochrones without respect to age is
an overly simplistic approach to characterizing the location of brown
dwarfs in the HR diagram.
The DM97 models reproduce the evolution in the brown dwarf isochones
especially well, however the absolute ages are anomalously young.

\section{DISCUSSION}

Converting the luminosity proxy into a consistent and accurate tool
for age measurements requires solving two problems:  (1)
establishing whether the ages inferred from luminosities are
consistent across a temperature range, and (2) comparing the
model-dependent age estimates with independent age estimates to
establish an absolute and correct age scale.

The empirical isochrone fits presented here lead to a self-consistent
luminosity ordering for nearby young associations.   Absolute ages are then estimated from
comparisons to isochrones for model tracks, but these comparisons lead to a wide range of age
estimates for different model tracks and within the same models for
stars of different temperatures. 
If uncorrected, these
discrepancies lead to significant spectral type (mass) dependences in
ages of young stars.  

The temperature dependency of the inferred ages arises from the mismatch
between empirical and theoretical slopes in $\log L$ vs temperature.
The source for this mismatch may be the evolutionary models,
although spectral type/temperature mismatches between different
methodologies and uncertainties in the spectral type/temperature
conversion also likely contribute to these discrepancies.  The BHAC-2015 models improved the
prescription for convection, while 
the Feiden models improve upon the
thermal structure in the atmosphere.  Both
the BHAC-2015 and Feiden models significantly improve the match between
predicted and observed stellar loci.  
Alternately, the PARSEC models apply an {\it ad hoc} change to the T-$\tau$
relationship to fit the color constraints of old clusters, and similar
tweaks may be able to reproduce loci of low mass young stars.

Ultimately, despite recent advances, some model improvements are still
necessary so that the evolution of
radius and temperature better reflects the observational constraints
of young stellar associations.  
Missing physics in the evolutionary models, including magnetic fields
and spots, may lead to mass dependent discrepancies or absolute errors
in age dating young clusters.
Age comparisons between clusters are self-consistent only when compared over
specific temperature ranges where the model and empirical isochrones
match.    The BHAC-2015 and Feiden models currently offer the best set of
pre-main sequence tracks for relative age estimates, although
mismatches at $<3500$ K still lead to temperature-dependent ages.
The empirical isochrones derived in this
paper offer a method to place stars of different temperatures on the
same age scale, albeit with some uncertainty introduced by the
possible temporal evolution of the isochronal slopes.

Table~\ref{tab:ages.tab} lists sets of ages for each cluster based on a
consistent application of a single model family at a single stellar
temperature.  The ages are selected from models that produce young,
intermediate, and old ages relative to the average model.
 Uncertainties in these ages are calculated from
the statistical uncertainties in the fits, as listed in
Table~\ref{tab:lumorder.tab}$^9$, and an estimate for the systematic
uncertainty in the median distance of the $\eta$ Cha, $\epsilon$ Cha,
and Upper Sco OB Associations.  
\footnotetext[9]{The luminosity errors in dex are asymmetric in
  linear space.  The listed errors are the median error in Myr.}

The reasonableness of these isochronal ages are testable by comparison to Li
depletion boundary ages$^{10}$.   The Tuc-Hor Moving Group has a Li
depletion boundary age of $41\pm2$ Myr from the BCAH98 evolutionary
models and $38\pm2$ Myr from the DM97 models \citep{Kraus2014}, so we
adopt $40\pm4$ Myr.  The $\beta$ Pic Moving Group has Li depletion
boundary ages of $21\pm4$ Myr from \citep{Binks2014} and $26\pm3$ Myr from the
Feiden models \citep{Malo2014b}.  We adopt an
Li depletion boundary age of $24\pm4$ Myr for the $\beta$ Pic Moving Group. 
\footnotetext[10]{The Li depletion boundary ages may be systematically overestimated by a
  few Myr if the radius dispersion within a cluster decreases the
  timescales for Li depletion \citep{Somers2014}.  On the other hand,
  inclusion of magnetic spots may lead to 30\% and 20\% underestimates
of the ages of the $\beta$ Pic and Tuc-Hor Moving Groups, respectively
\citep{Jackson2014}.}

\begin{figure*}[!t]
\epsscale{1.}
\plottwo{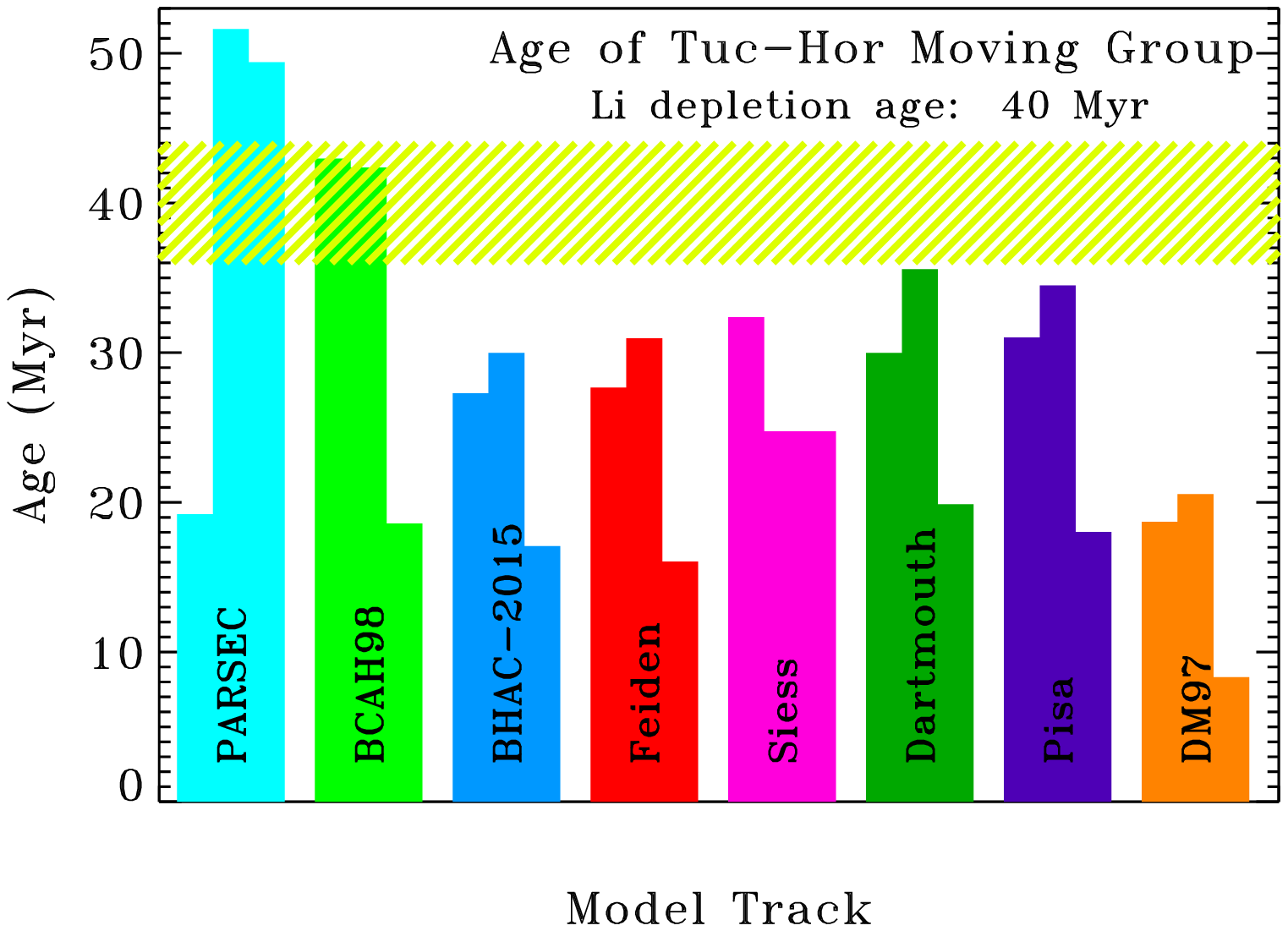}{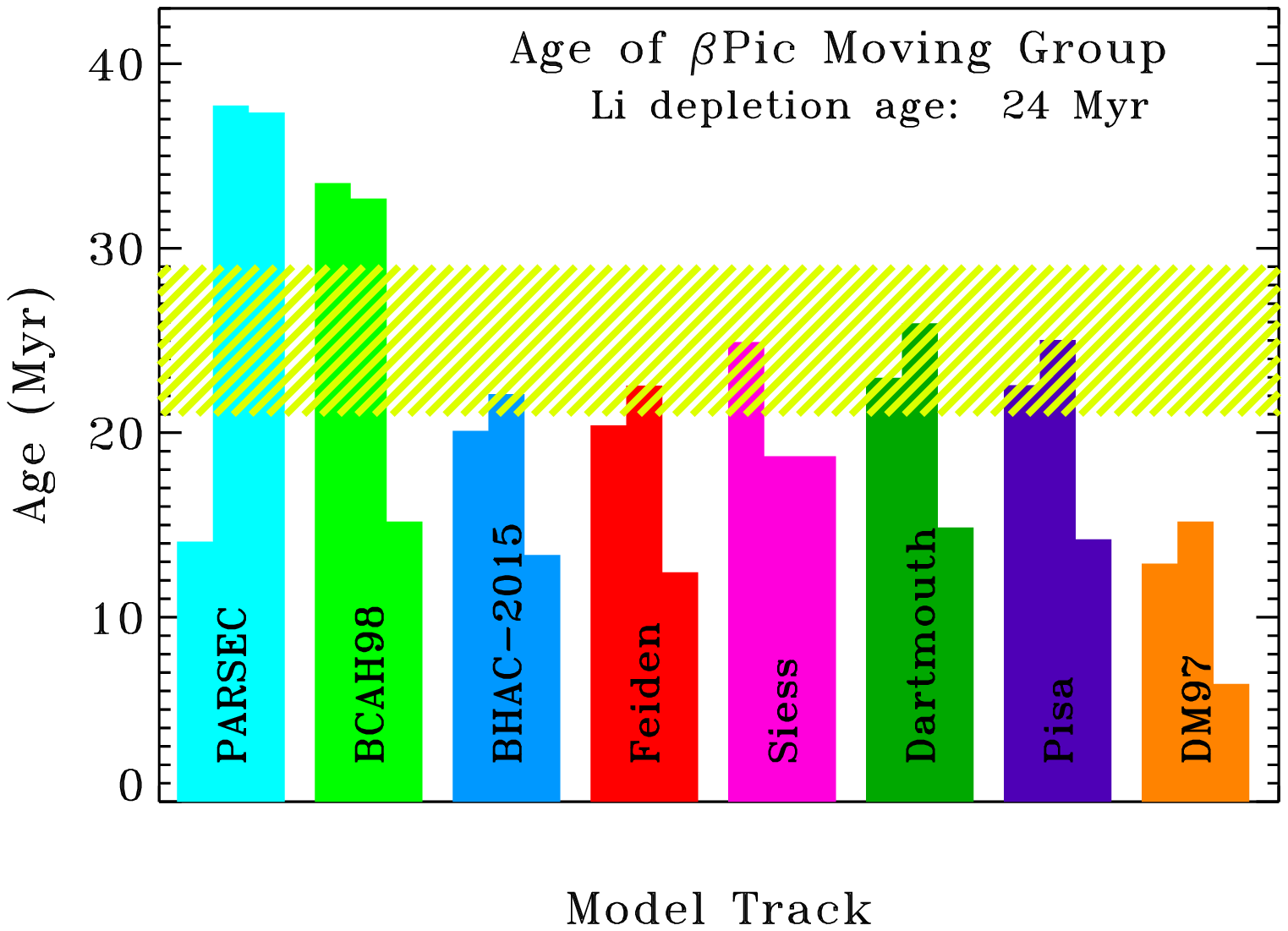}
\plottwo{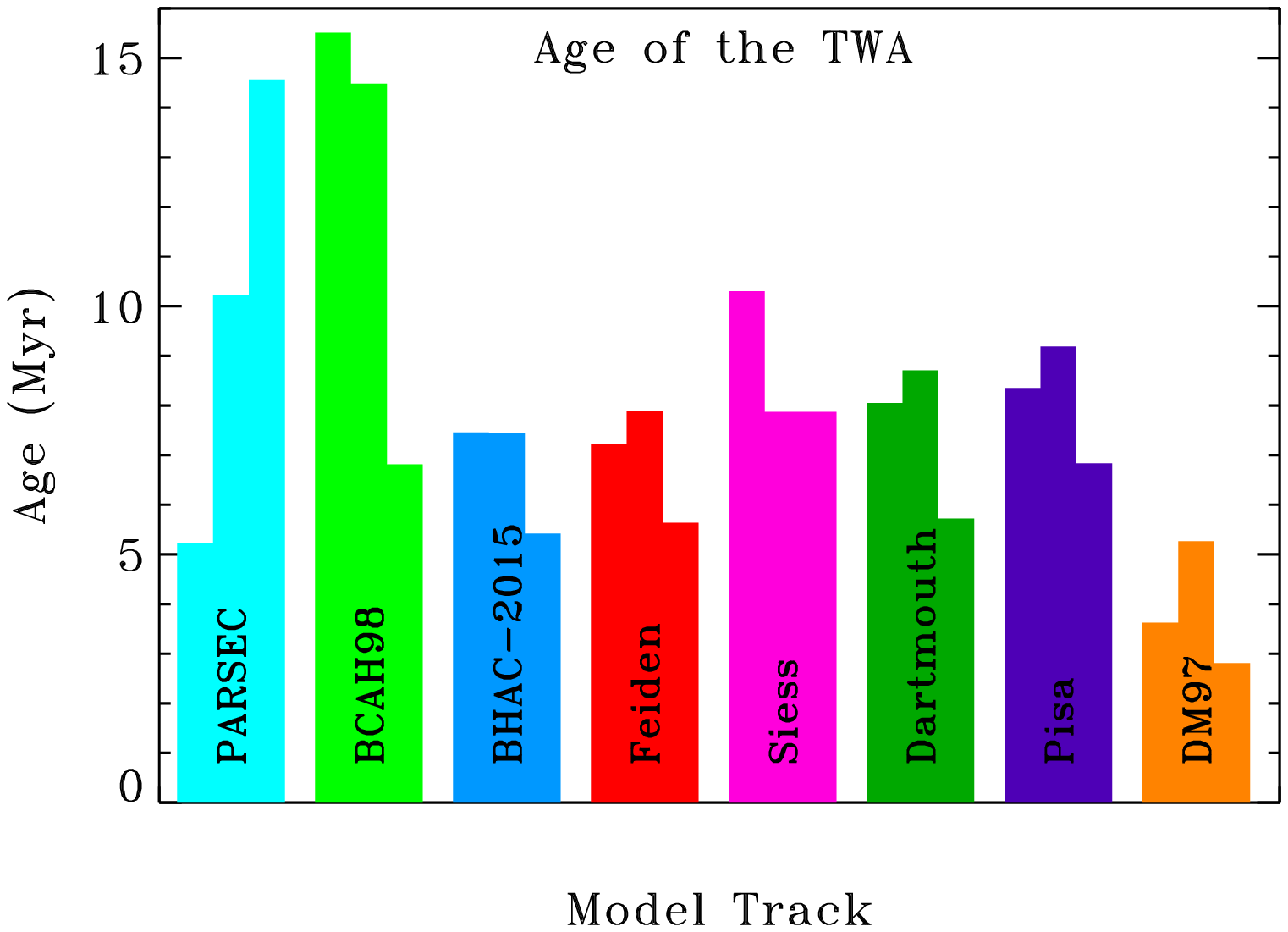}{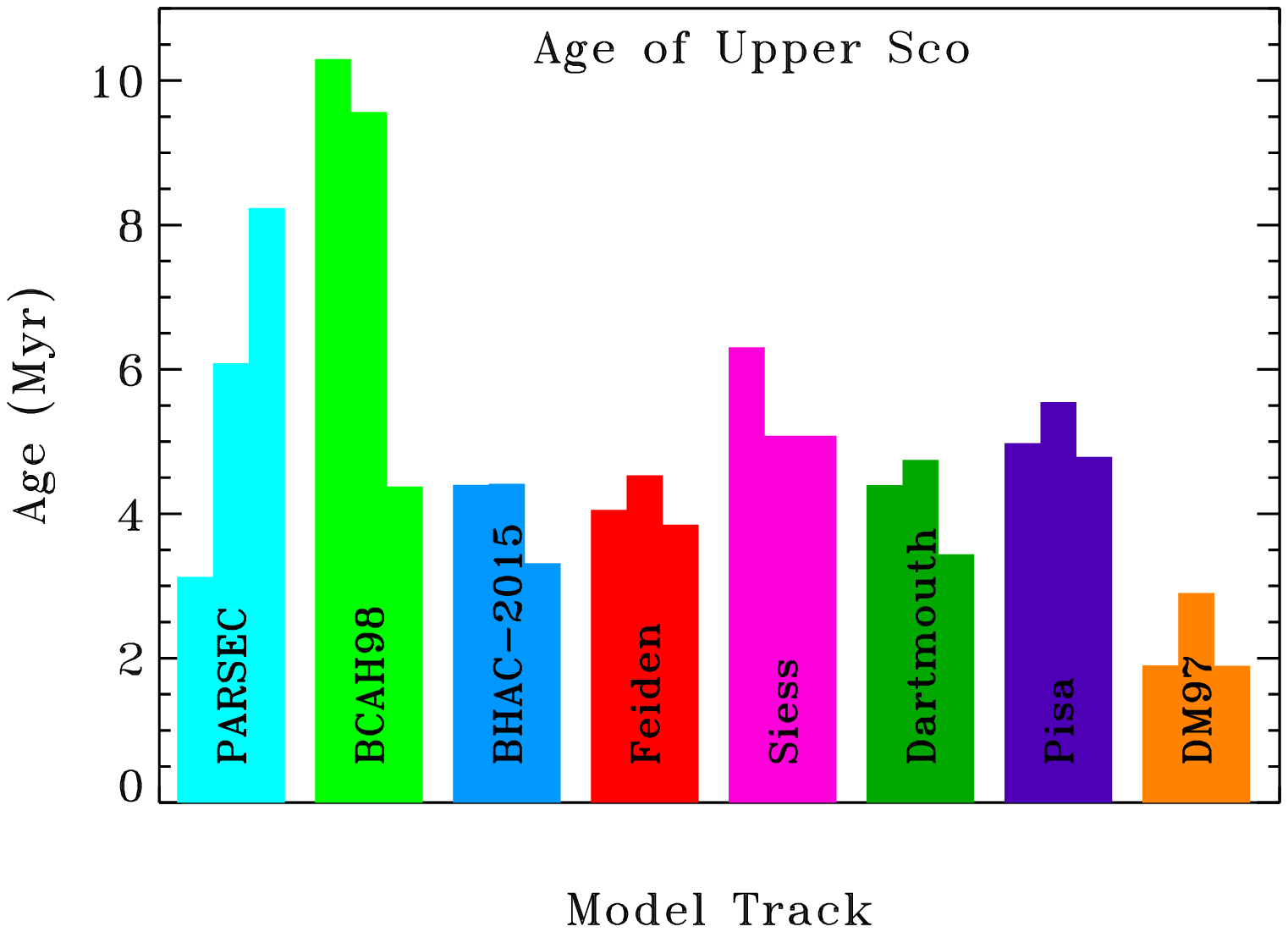}
\caption{Age estimates of Tuc Hor Moving Group, $\beta$ Pic Moving
  Group, the TW Hya Association, and Upper Sco OB Association, as calculated from eight different
  models for 4200 K, 3800 K and 3400 K stars (from left to right
  within each model).  The horizontal hatched
  yellow region shows the Li
  depletion boundary ages of $24\pm4$ and $40\pm4$ Myr$^6$ for the $\beta$ Pic and Tuc-Hor Moving Groups 
  \citep{Binks2014,Malo2014b,Kraus2014}.}
\label{fig:ages_hist}
\vspace{5mm}
\end{figure*}

Figure~\ref{fig:ages_hist} and Table~\ref{tab:ages.tab} compares the
Li depletion boundary ages to the isochronal ages
estimated here from the different model tracks.  No single model of 
low mass stars
reproduces the Li depletion boundary age for both the Tuc-Hor and
$\beta$ Pic Moving Groups.  Most models underpredict the age of the
Tuc-Hor Moving Group by $\sim 10$ Myr (or 0.1 dex).  On the other
hand, most models reproduce the age of the $\beta$ Pic Moving Group.
The DM97 models yield especially
young ages, likely because the isochrones of low-mass stars have
temperatures that are initially too high (see description in \S2.5).  
 Both the Feiden and BHAC-2015 models have similar isochrones that
  yield remarkably similar ages.

When applied consistently, the Li depletion boundary ages for the 
$\beta$ Pic Moving Group implies a $\sim 5$ Myr old age for low mass
stars in the Upper Sco OB Association,
younger than the 11 Myr age measured from intermediate and high mass stars by
\citet{Pecaut2012}.    The 11 Myr age of Upper Sco would require ages
of $\sim 40$ and 55 Myr for the $\beta$ Pic and Tuc-Hor Moving Groups.
Either the intermediate mass or low mass age (based on Li depletion
boundary ages of older associations) of the Upper Sco OB
Association is incorrect, or the identification and evaluation of the
low mass and/or higher mass stars is biased in some way.

Errors in ages may be caused by either errors in the
birthline or in the contraction rates.  A bulk subtraction of 5 Myr to
intermediate mass isochrones would shift the intermediate mass
isochrones sufficiently for Upper Sco and would not cause any significant
discrepancy with the Li depletion age of the $\beta$ Pic Moving
Group.  The age estimate for the $\beta$ Pic Moving Group from late F
and G stars \citep{Mamajek2014} would
also not change significantly because these stars are preferentially
lower mass than the intermediate/high mass stars that led to the old age
estimate for the Upper Sco OB Association.
An addition of 5 Myr to low-mass isochrones would also lead to
consistent ages, but would encounter severe problems by forcing the youngest
known clusters to have ages of 5
Myr.  Alternately, the contraction rates for either low-mass or intermediate mass stars
may be in error.  For low mass stars, the inclusion of magnetic fields
and spots in evolutionary models will affect contraction rates.  For
intermediate and high mass stars, the inclusion of rotation 
leads to older ages for intermediate and high-mass stars.  

Observational errors relative to the benchmark associations may also lead to an underestimate
of the age.  A bulk increase in temperatures by 250 K would lead to an
age of 10 Myr, but such a large systematic difference in spectral
typing is unlikely.  Alternately, the luminosities could be
overestimated by 0.2 dex if the distances are biased and overestimated
by 25\%.  Assuming $A_V=0$ mag, which disagrees with the colors of
Upper Sco stars, would lead to only a 0.07 dex increase in the
luminosities.

\section{Conclusions}

In this paper, we calculate empirical isochrones in HR diagrams for low mass stars
and brown dwarfs in young ($5-40$ Myr) nearby associations.   The sample consists of stars
that have accurate
distances, low extinction, and are not affected by the presence of
disks or accretion.
The simple formulation of
the empirical isochrone assumes that two slopes, one applied to stars
with temperatures 3200/3400--5000 K and one for stars between 2800--3200/3400 K, applies to
the stellar locus of all young associations.  This choice leads to reasonable fits to
the five young benchmark clusters and to the stellar and brown dwarf
locus of the Upper Sco OB Association.

A successful theory of
pre-main sequence evolution should match these empirical fits.
The Dartmouth, Pisa, and BCAH98-1.9 pre-main sequence isochrones
between 3200--5500 are consistent with the measured slopes.  The
slopes from DM97, BCAH98, and Siess isochrones are steeper than
observed, so that very low mass stars will be assessed as
younger than solar mass stars.  The PARSEC isochrones have the
opposite behavior, so that solar mass stars are assessed as younger than very
low-mass stars.  The PARSEC and BCAH98 models typically lead to
the oldest ages, while the ages obtained from the DM97 models are
anomalously young.   Advances in prescriptions for convection and
  the thermal structure of the atmosphere in the BHAC-2015 and
  Feiden models, respectively, have improved the match to empirical stellar loci at
$>3500$ K and match the decrease in luminosity at $<3500$ K, though
with a temperature offset relative to established spectral
type-temperature conversions. 

Applying the empirical isochrone to associations provides a
statistically robust age ordering, which is independent of spectral
type.  Our age ordering is based on the 3200--5000 K region.  
The assumption that the 2800--3200 K isochrones have the same slope
for both the young and old associations is
likely overly simplistic.  Steeper slopes at younger ages, especially
for the Upper Sco OB Association, instead
indicates some evolution in the brown dwarf locus.  Because the
relationship between luminosity and temperature for brown dwarfs is
steep, any use of brown dwarfs in age estimates requires accurate
temperatures and should be used with caution.

Fitting these isochrones to the six associations discussed in this
paper leads to a luminosity/age ordering mostly as expected from previous
results, with the Upper Sco ($L_{{\rm 4200}}/L_\odot=-0.23$, $\eta$
Cha ($-0.28$), and $\epsilon$ Cha ($-0.32$), 
Associations roughly co-eval, followed by the TW Hya Association
($-0.40$) and
the $\beta$ Pic ($-0.65$ and Tuc-Hor ($-0.72$) Moving Groups.  For any
given association, age estimates of different models evaluated at different temperatures span
0.8 dex.   As a consequence, comparisons of ages between different
associations should consistently apply a single luminosity-age scaling.
The BHAC-2015 and Feiden pre-main sequence models yield very similar
ages for low-mass stars.  
Ages can be adopted to be consistent with
the Li depletion boundary age of the $\beta$ Pic Moving Group, but
these models 
imply a $\sim 5$ Myr age for the Upper Sco OB
assocation for low mass stars, younger than the 11 Myr age obtained from
intermediate mass and high mass stars stars.

\acknowledgements

We thank the referee, Cameron Bell, for a careful read of the manuscript,
which led to substantial improvements in the robustness and clarity of
results and in the self-consistency within the paper.
GJH thanks Adam Kraus for discussions on preparing the paper and Jessy
Jose for valuable comments on the paper.  GJH is also grateful for the role of the 2014
Oort Workshop on Episodic Accretion, 
jointly organized by Ewine van Dishoeck and Neal Evans, in motivating
this analysis, and discussions with Lee Hartmann and
Isabelle Baraffe at the workshop on early ideas for this paper.
We appreciate Gregory Feiden sharing the new Dartmouth
tracks prior to their publication and for discussion of those tracks
and magnetic fields.   GJH is supported by a Youth Qianren grant and general
grant \# 11473005 awarded by the National Science Foundation of China.
This research has made use of the VizieR catalogue access tool, CDS, Strasbourg, France.

\section{Appendix A:  Target photometry and bolometric corrections}

In \S 2.2--2.3, we describe the collection and use of photometry to
identify disk sources, to tweak bolometric corrections calculated from
models, and to measure extinctions for members of the Upper Sco OB
Association.  This appendix describes how the photometry is
implemented in this paper.

Filter curves and zero points are obtained from the Virtual
Observatory Filter Profile Service$^{11}$.  The BT-Settl models were
convolved with the filter transmission profiles in photon space, which
may lead to minor offsets relative to integrations in energy space
\citep{Cohen2003}.  The number of counts is then converted into a
magnitude from the filter zero point function.
Synthetic photometry is then calculated from the zero-point flux for
the filter and the central wavelength of the filter.  The listed central
wavelengths are the average wavelength of the filter, calculated by
integrating the transmission over the filter bandpass in wavelength
space.
\footnotetext[11]{http://voservices.net/filter/}

\begin{figure*}[!t]
\epsscale{1.}
\plottwo{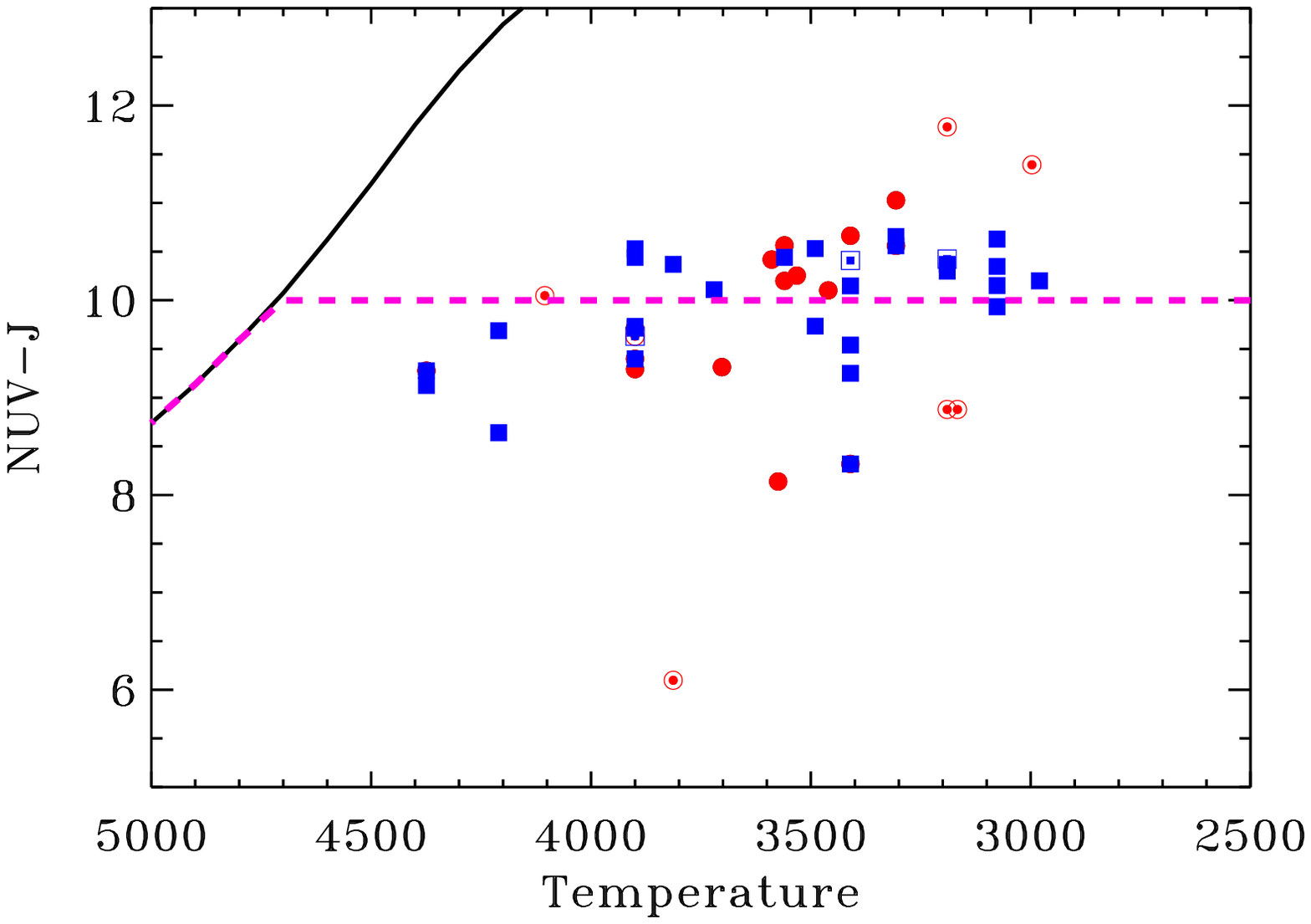}{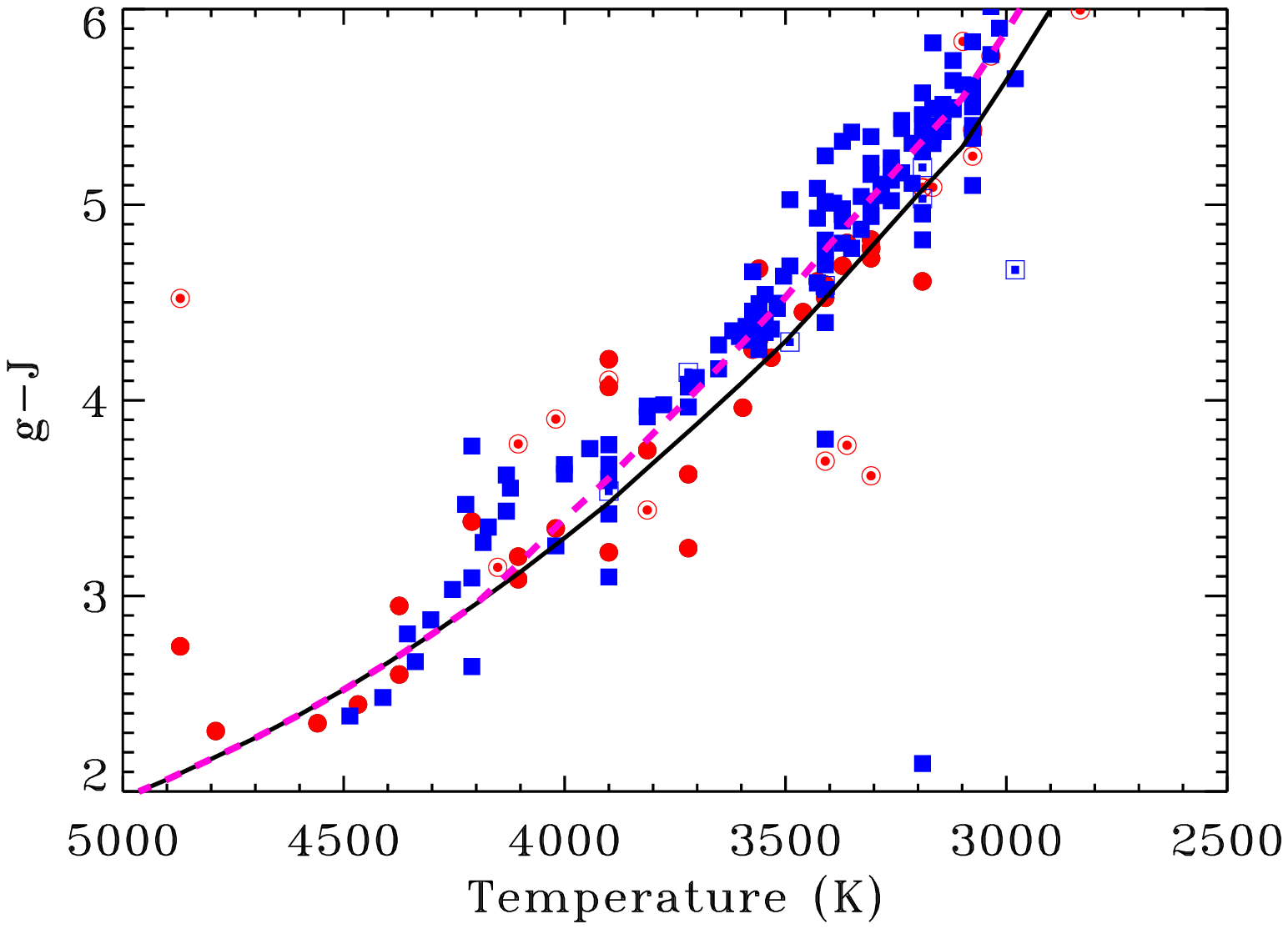}
\plottwo{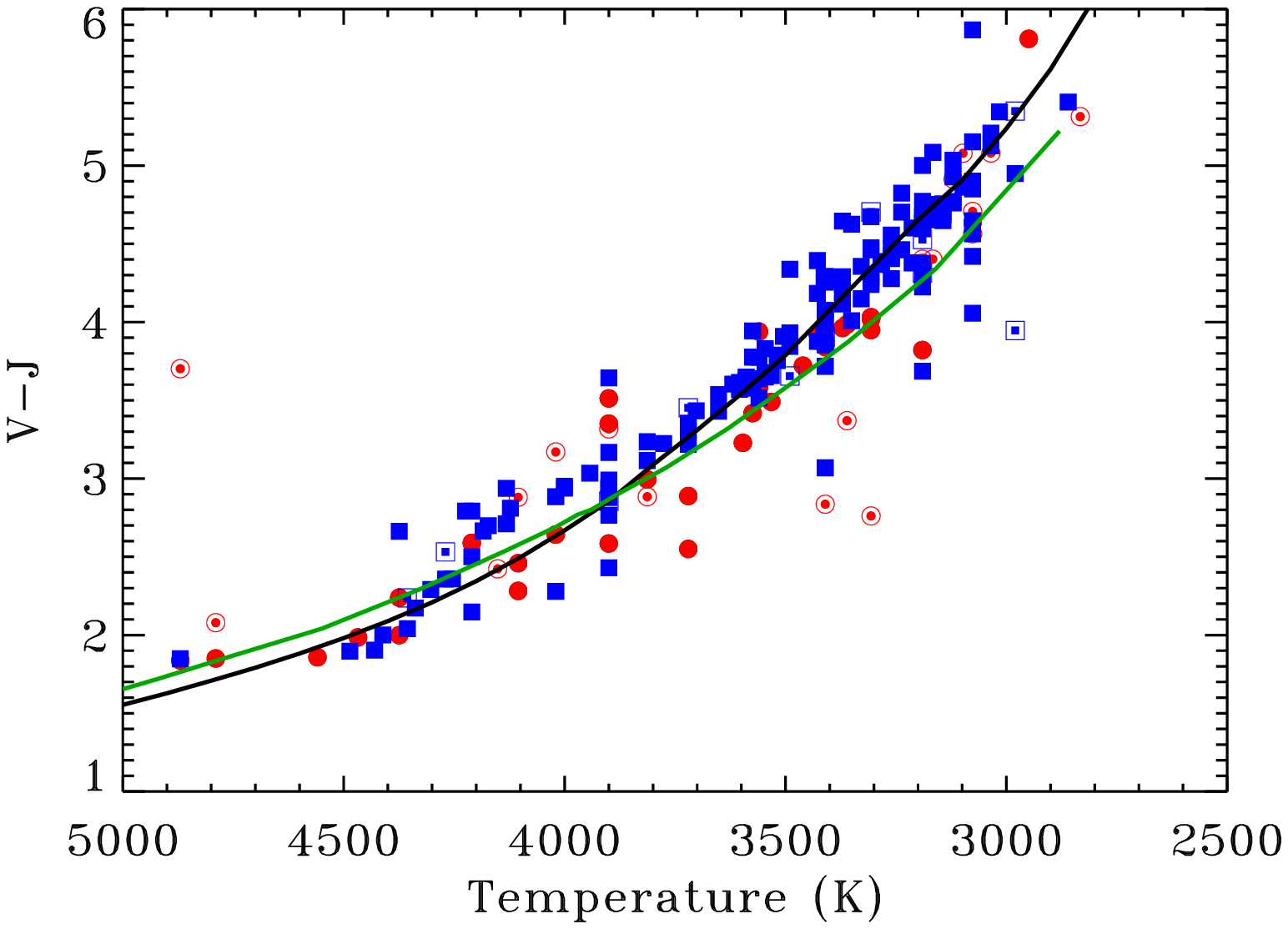}{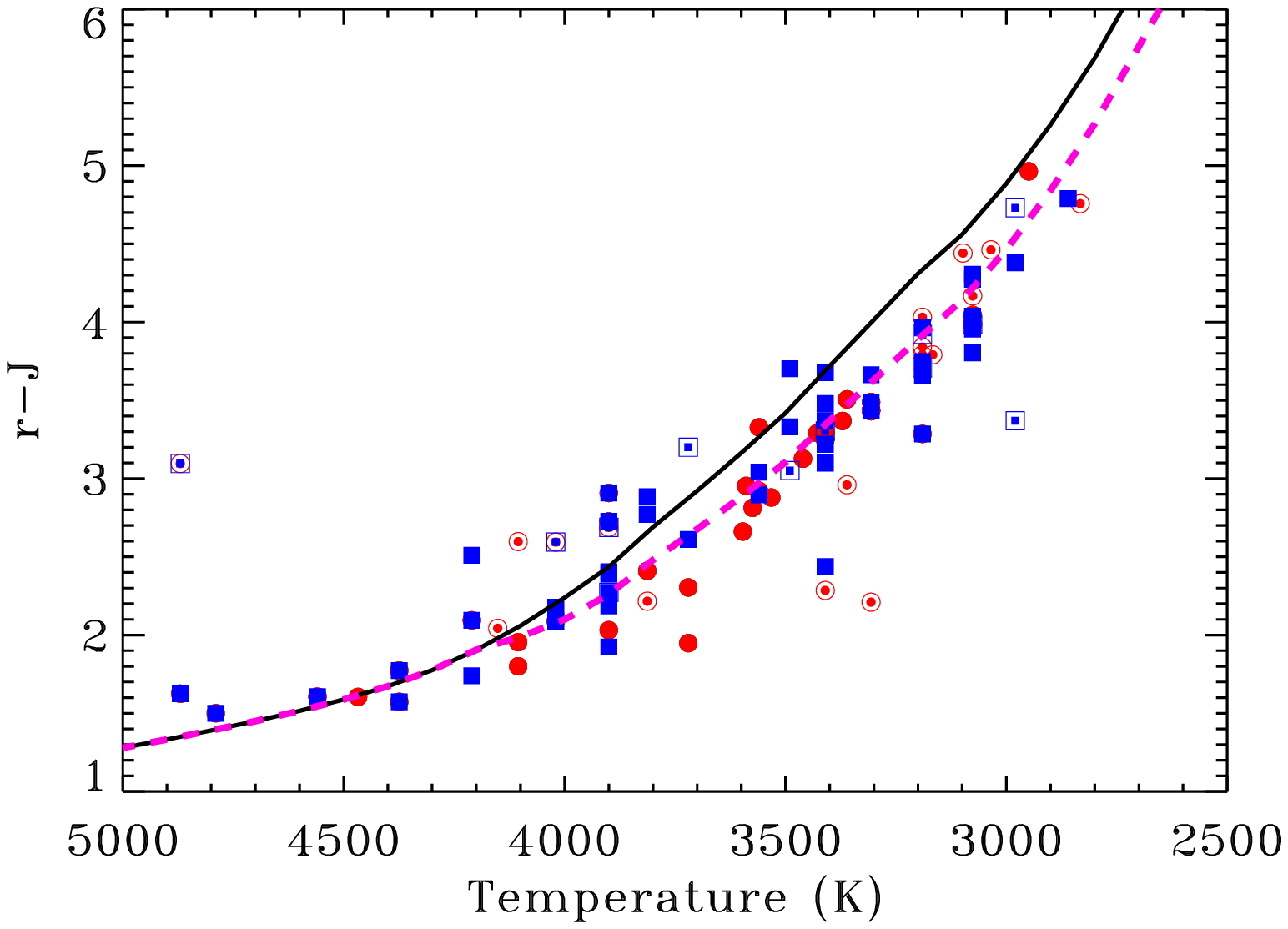}
\plottwo{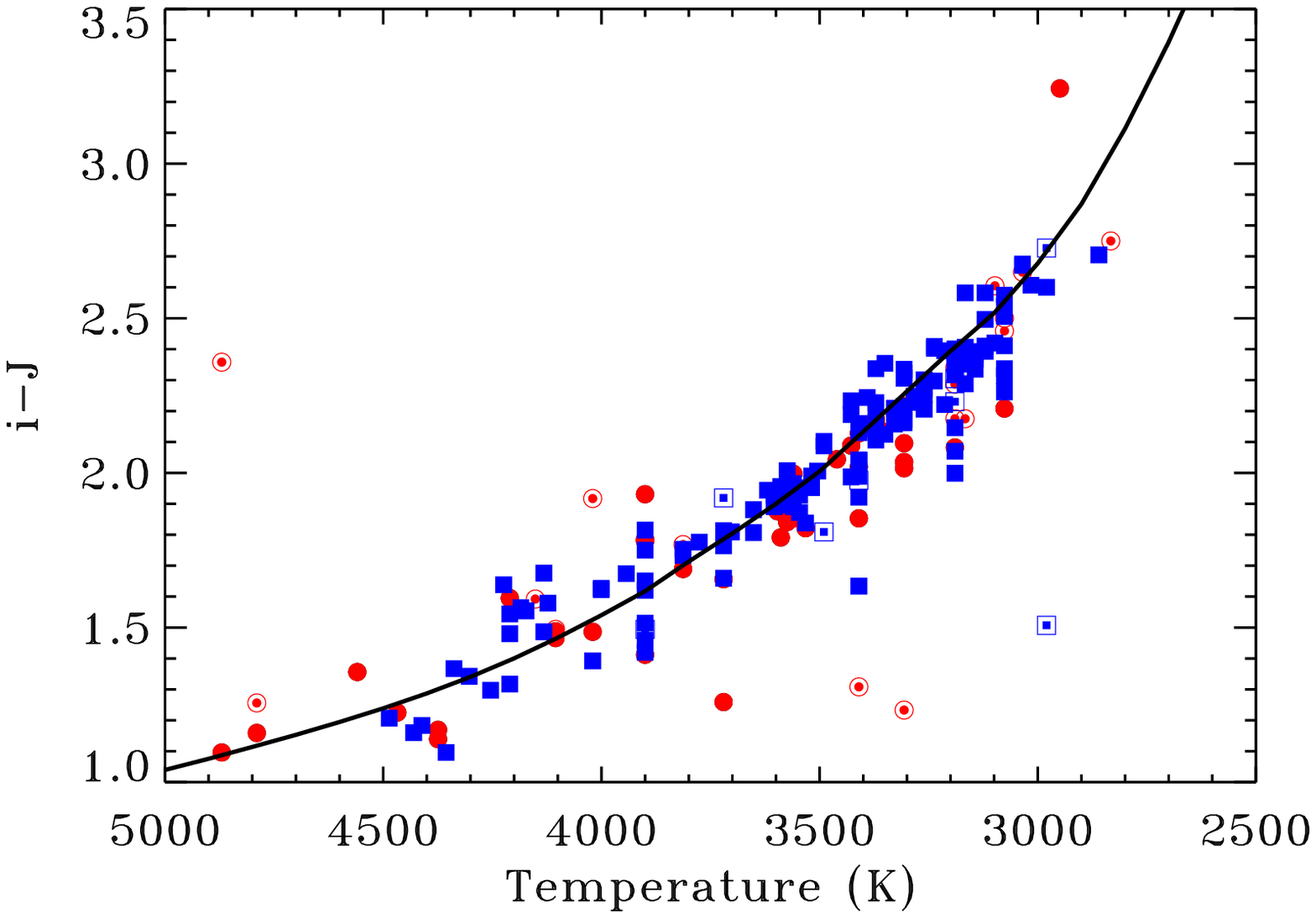}{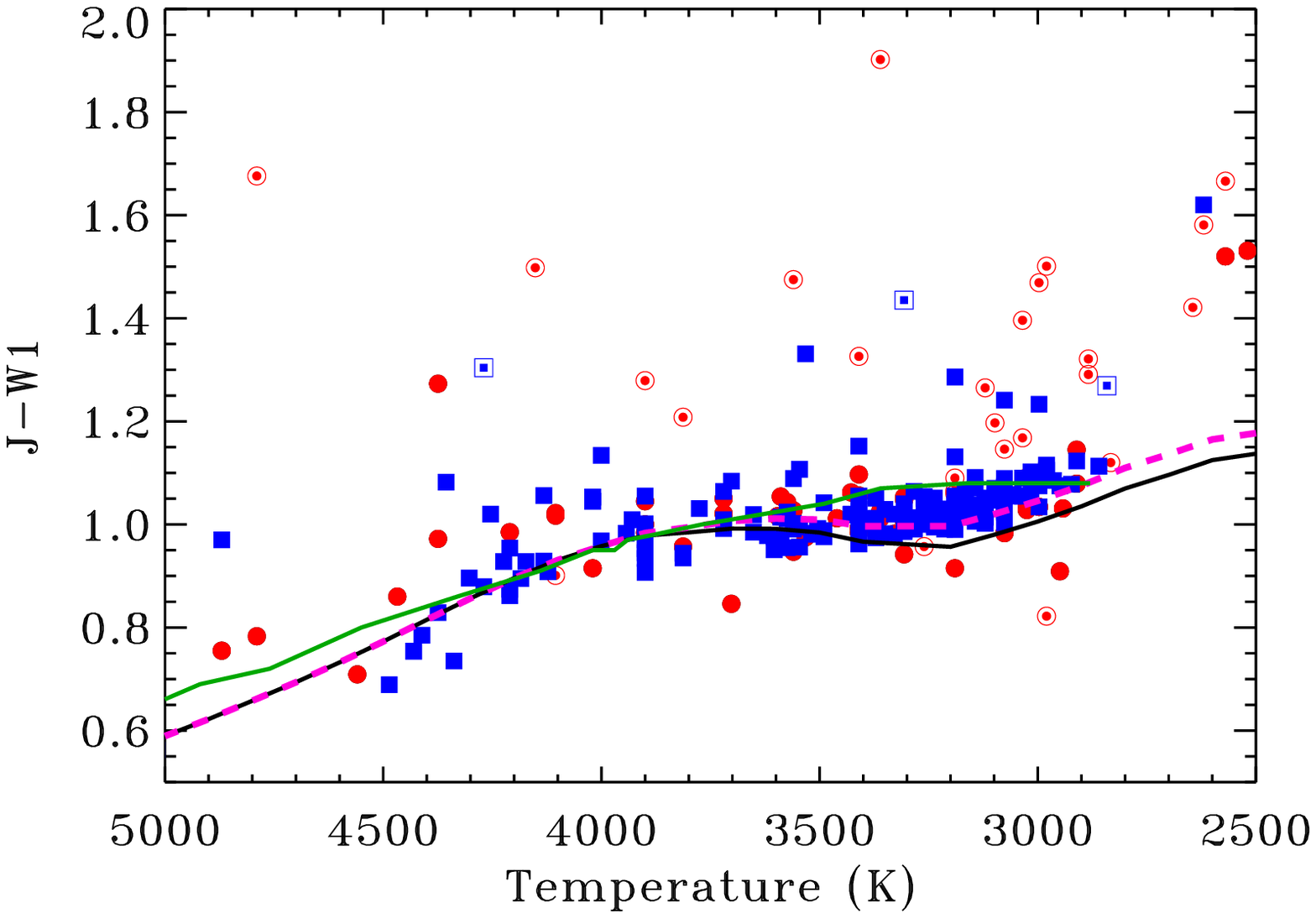}
\plottwo{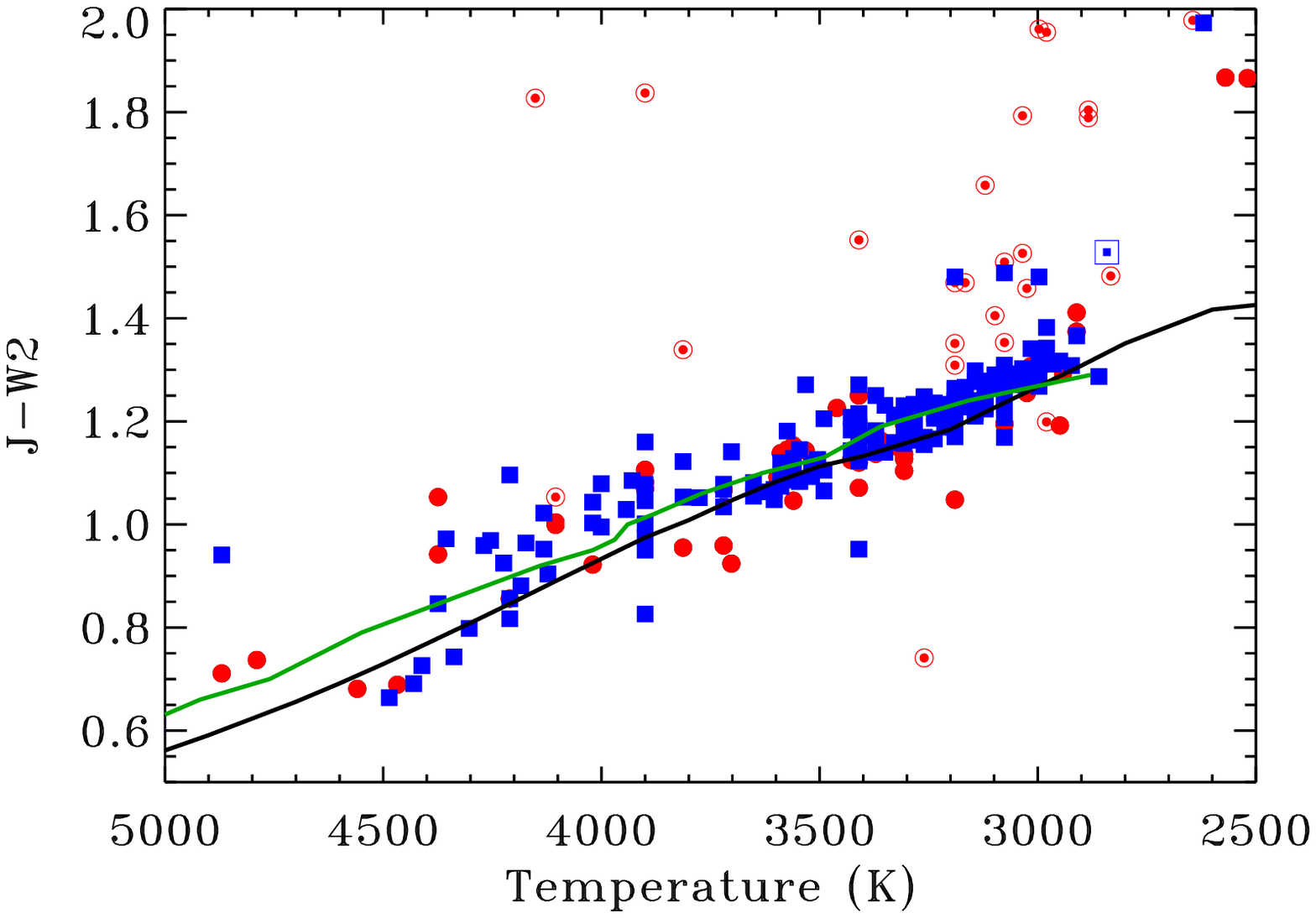}{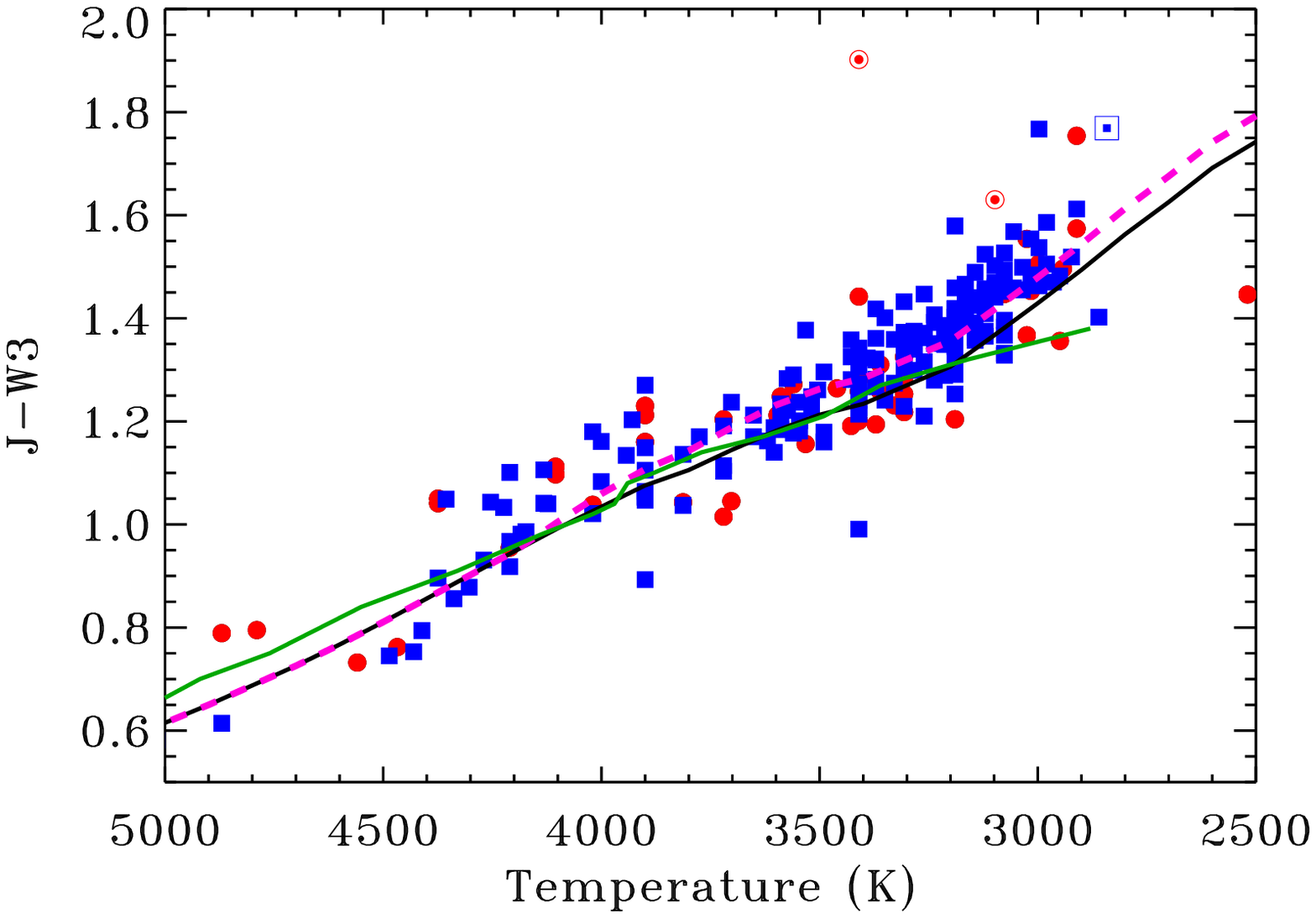}
\caption{Color-temperature plots for members of young (red circles) and old
  (blue squares) nearby associations.  All colors include J to
  evaluate bolometric corrections from the J band.  Synthetic colors
  from BT-Settl models are plotted as the black line, while colors for
young (5--30 Myr) stars from \citet{Pecaut2013} are plotted in green.
In cases where the observed colors deviate from the model colors, the
colors adopted for the bolometric correction calculations are shown as
a dashed purple line.  Objects with disks, identified from a K-W2 or K-W3
excess (Fig.~\ref{fig:diskcolor}), are shown here as the small filled circles or
squares surrounded by an outer circle or square.}
\label{fig:colors}
\end{figure*}

Figure~\ref{fig:colors} shows temperature-color plots that compare the
observed photometry to synthetic photometry from the BT-Settl models
and, when possible, to photometry of \citet{Pecaut2013}.  In cases
where the observed and synthetic colors (always relative to J) are mismatched, the model
fluxes at the corresponding wavelength are corrected to reproduce the observed
colors.  The J-band flux is always kept constant, so all corrections
are relative to the J-band.
The bolometric corrections are then calculated from the
synthetic J-band photometry and the integrated emission in the full
model spectrum.

Finally, the absolute bolometric magnitude $M_{bol}$ is calculated from
the distance-corrected J-band magnitude, $M_J$, and the bolometric
correction, BC, $M_{bol}=M_J+BC$.  The zero point for $M_{bol}$
($3.055\times10^{35}$ erg cm$^{-2}$ s$^{-1}$) and
the assumed $L_\odot=3.83\times10^{33}$ erg cm$^{-2}$ s$^{-1}$ are obtained from the 1999 IAU Zero Point Scale,
as discussed at the webpage of Eric Mamajek listing Basic Astronomical
Data for the Sun$^{12}$.
\footnotetext[12]{https://sites.google.com/site/mamajeksstarnotes/basic-astronomical-data-for-the-sun}

\pagebreak

\end{CJK*}

\end{document}